\documentclass[aps,prc,letterpaper,11pt,twoside,tightenlines,nofootinbib,showpacs,preprint]{revtex4}
\usepackage{graphicx}
\usepackage{epsfig,float}
\usepackage[sort&compress]{natbib}
\usepackage{subfigure}
\usepackage{amsmath}
\usepackage{amsfonts}
\usepackage{cancel}
\usepackage{lmodern,dsfont}
\usepackage{amssymb}
\usepackage{hyperref}
\usepackage{multirow}

\usepackage{color}

\begin{document}
\arraycolsep1.5pt
\newcommand{\Ima}{\textrm{Im}}
\newcommand{\Rea}{\textrm{Re}}
\newcommand{\mev}{\textrm{ MeV}}
\newcommand{\gev}{\textrm{ GeV}}
\newcommand{\dtres}{d^{\hspace{0.1mm} 3}\hspace{-0.5mm}}
\newcommand{\rts}{ \sqrt s}
\newcommand{\non}{\nonumber \\[2mm]}
\newcommand{\eps}{\epsilon}
\newcommand{\half}{\frac{1}{2}}
\newcommand{\thalf}{\textstyle \frac{1}{2}}
\newcommand{\Nmass}{M_{N}} % mass of nucleon
\newcommand{\delmass}{M_{\Delta}} % mass of delta
\newcommand{\pimass}{\mu}  % mass of pion 
\newcommand{\rhomass}{m_\rho} % mass of rho
\newcommand{\piNN}{f}      % coupling of pi NN
\newcommand{\rhocoup}{g_\rho} % universal coupling to rho
\newcommand{\fpi}{f_\pi} % pion decay constant fpi
\newcommand{\f}{f} % pion decay constant fpi
\newcommand{\nucfld}{\psi_N} % nucleon field
\newcommand{\delfld}{\psi_\Delta} % delta field
\newcommand{\fpiNN}{f_{\pi N N}} % coupling of pi N N 
\newcommand{\fpiND}{f_{\pi N \Delta}} % coupling of pi N delta 
\newcommand{\GMquark}{G^M_{(q)}} % magnetic coupling for quark 
\newcommand{\vecpi}{\vec \pi}
\newcommand{\vectau}{\vec \tau}
\newcommand{\vecrho}{\vec \rho}
\newcommand{\delmu}{\partial_\mu}
\newcommand{\delMu}{\partial^\mu}
\newcommand{\nn}{\nonumber}
\newcommand{\bi}{\bibitem}
\newcommand{\vs}{\vspace{-0.20cm}}
\newcommand{\be}{\begin{equation}}
\newcommand{\ee}{\end{equation}}
\newcommand{\ba}{\begin{eqnarray}}
\newcommand{\ea}{\end{eqnarray}}
\newcommand{\ropi}{$\rho \rightarrow \pi^{0} \pi^{0}
\gamma$ }
\newcommand{\roeta}{$\rho \rightarrow \pi^{0} \eta
\gamma$ }
\newcommand{\omepi}{$\omega \rightarrow \pi^{0} \pi^{0}
\gamma$ }
\newcommand{\omeeta}{$\omega \rightarrow \pi^{0} \eta
\gamma$ }
\newcommand{\ul}{\underline}
\newcommand{\del}{\partial}
\newcommand{\rth}{\frac{1}{\sqrt{3}}}
\newcommand{\rsix}{\frac{1}{\sqrt{6}}}
\newcommand{\sq}{\sqrt}
\newcommand{\fr}{\frac}
\newcommand{\pr}{^\prime}
\newcommand{\ov}{\overline}
\newcommand{\Gm}{\Gamma}
\newcommand{\rw}{\rightarrow}
\newcommand{\rgl}{\rangle}
\newcommand{\De}{\Delta}
\newcommand{\Dp}{\Delta^+}
\newcommand{\Dm}{\Delta^-}
\newcommand{\Dz}{\Delta^0}
\newcommand{\Dpp}{\Delta^{++}}
\newcommand{\Sg}{\Sigma^*}
\newcommand{\Sp}{\Sigma^{*+}}
\newcommand{\Sm}{\Sigma^{*-}}
\newcommand{\Sz}{\Sigma^{*0}}
\newcommand{\X}{\Xi^*}
\newcommand{\Xm}{\Xi^{*-}}
\newcommand{\Xz}{\Xi^{*0}}
\newcommand{\Om}{\Omega}
\newcommand{\Omm}{\Omega^-}
\newcommand{\kp}{K^+}
\newcommand{\kz}{K^0}
\newcommand{\pip}{\pi^+}
\newcommand{\pim}{\pi^-}
\newcommand{\piz}{\pi^0}
\newcommand{\et}{\eta}
\newcommand{\kb}{\ov K}
\newcommand{\km}{K^-}
\newcommand{\kbz}{\ov K^0}
\newcommand{\ksb}{\ov {K^*}}

\newcommand{\red}{\textcolor{red}}
\newcommand{\blue}{\textcolor{blue}}

\def\tstrut{\vrule height2.5ex depth0pt width0pt} % used in tables
\def\jtstrut{\vrule height5ex depth0pt width0pt} % used in tables

\title{The width of the $\omega$ meson in the nuclear medium}

\author{A. Ramos$^{1}$, L. Tolos$^{2,3}$, R. Molina$^4$ and E. Oset$^5$}
\affiliation{
$^1$ Departament d'Estructura i Constituents de la Mat\`eria and Institut de
Ci\`{e}ncies del Cosmos, Universitat de Barcelona, Mart\'{\i} i Franqu\`es 1, E-08028
Barcelona, Spain\\
$^2$ Instituto de Ciencias del Espacio (IEEC/CSIC) Campus Universitat Aut\`onoma
de Barcelona, Facultat de Ci\`encies, Torre C5, E-08193 Bellaterra (Barcelona), 
Spain\\
$^3$ Frankfurt Institute for Advanced Studies (FIAS). Johann Wolfgang Goethe University. Ruth-Moufang-Str. 1. 60438 Frankfurt am Main. Germany\\
$^4$Research Center for Nuclear Physics (RCNP),
Mihogaoka 10-1, Ibaraki 567-0047, Japan\\
$^{5}$Departamento de F\'{\i}sica Te\'orica and IFIC, Centro Mixto Universidad de 
Valencia-CSIC,
Institutos de Investigaci\'on de Paterna, Aptdo. 22085, E-46071 Valencia,
Spain
}

\date{\today}

\begin{abstract}

We evaluate the width of the $\omega$ meson in nuclear matter.
We consider the free decay mode of the $\omega$ into three pions, which
is dominated by $\rho \pi$ decay, and replace the $\rho$ and
$\pi$ propagators by their medium modified ones. We also take into
account the quasielastic and inelastic processes induced by a vector-baryon
interaction dominated by vector meson exchange,
as well as the
contributions coming from the $\omega \to K \bar K$ mechanism with medium
modified
$K$, $\bar K$ propagators.
 We obtain a substantial
increase of the $\omega$ width in the medium, reaching a value of $129 \pm 10$
MeV at normal nuclear matter density for an $\omega$ at rest, which comes mainly from $\omega N
\to \pi \pi N, \omega N N \to \pi NN$ processes associated to the dominant
$\omega \to \rho\pi$ decay mode. The value of the width increases moderately
with momentum, reaching values of around 200 MeV at 600~MeV/c.
%No substantial shift of the mass is found from the calculations. ?????????? 
\end{abstract}

\maketitle

\section{Introduction}
\label{Intro} 
The interaction of vector mesons with nuclei is a subject that has attracted
much attention. The large amount of work, experimental and theoretical, has
been reviewed extensively in \cite{rapp} and more recently in 
\cite{Hayano:2008vn,Leupold:2009kz}. Another recent review \cite{review},
reporting on progress done using the local hidden gauge theory for the
interaction of vectors \cite{hidden1,hidden2,hidden4,Harada:2003jx}, has brought a new
perspective into this topic. 

Although originally there were some hopes that the mass of the vector mesons
would be drastically reduced in the nuclear medium, devoted studies concluded
that this was not the case
\cite{Rapp:1997fs,Peters:1997va,Rapp:1999ej,Urban:1999im,Cabrera:2000dx,
Cabrera:2002hc,Cabrera:2003wb} and recent experiments have confirmed the results
of these
calculations \cite{Hayano:2008vn,Leupold:2009kz,na60,Wood:2008ee}.

In addition to the emblematic $\rho$ meson, one of the vector mesons which has
been more thoroughly investigated experimentally is the $\omega$ meson. Apart
from investigations using heavy ions, other reactions, easier to interpret,
have been carried out using proton beams on nuclei at KEK
\cite{Ozawa:2000iw,Tabaru:2006az,Sakuma:2006xc} or photonuclear reactions
looking for dileptons in the final state \cite{Wood:2008ee,Wood:2010ei}.
However, one of the most detailed experiments on the $\omega$ properties in the
nucleus has been done at ELSA (Bonn), using photoproduction on nuclei and
detecting the $\omega$ through its $\pi^0 \gamma$ decay mode
\cite{Trnka:2005ey}. Originally, the analysis of the experimental results led
the authors of \cite{Trnka:2005ey} to claim the first observation of the
in-medium modifications of the $\omega$ meson mass, with a reduction of
about 100 MeV at normal nuclear matter density. Yet,
it was shown in \cite{Kaskulov:2006zc} that the results of \cite{Trnka:2005ey}
were tied to a particular choice of background. A
thorough study of the background processes in a revised analysis of the
experiment \cite{Nanova:2010sy} concluded that {the experiment was not sensitive
enough to possible changes of}
% there was nosignificant shift of 
the in-medium $\omega$ mass. {Preliminary results from the momentum
distribution appear to exclude shifts of the mass by 16 $\%$ \cite{trento}.}
    
  Simultaneously, the possibility that some signal reported in
\cite{tesistrnka,Metag:2007zz} could  indicate the formation of a $\omega$
meson bound state in nuclei was ruled out, showing that the double hump
structure observed in the experiment was due to uncorrelated $\pi^0 \gamma$
production events together with the $\omega$ production and subsequent
$\pi^0 \gamma$ decay, which scaled differently with growing nuclear mass
\cite{Kaskulov:2006fi}.  The expectations that the use of the mixed events method to
separate the background from the signal could solve the problem
\cite{Metag:2007hq} were also shown unproductive in \cite{mixed}, due to the
fast exponential fall of the mass distribution which made inapplicable this
otherwise successful tool. 
  
  In summary, the thorough experimental and theoretical work along these
reactions led to clarify the issue and set warnings for the analysis of similar
reactions used to extract medium modifications of vector meson properties.
  
  Yet, there is some physical information that  survived the close scrutiny of
the former works, and this is the large width of the $\omega$ in the medium
found in \cite{Kotulla:2008aa} and also studied in \cite{Kaskulov:2006zc}. 
The CBELSA/TAPS collaboration reported a width 
of 130-150 MeV at normal nuclear matter density \cite{Kotulla:2008aa}, while
the analysis based on a Monte Carlo simulation taking into account the possible
reactions in the experimental set-up in the vicinity of the $\omega$ meson 
produced a distribution compatible with a width
of the order of 100 MeV \cite{Kaskulov:2006zc}. The theoretical
understanding of this large width is a challenge that we take in the present
work.

  The theoretical determination of the $\omega$ properties in the medium has
been addressed in numerous studies
\cite{Jean:1993bq,Klingl:1997kf,Saito:1998wd,Tsushima:1998qw,Friman:1998fb,
Klingl:1998zj,Post:2000rf,Saito:1998ev,Lykasov:1998ma,Sibirtsev:2002de,
DuttMazumder:2000ys,Lutz:2001mi,Zschocke:2002mp,DuttMazumder:2002me,
Riek:2004kx,Muhlich:2003tj,Eichstaedt:2007zp,Muehlich:2006nn,Muhlich:2006ps,
Steinmueller:2006id,
Martell:2004gt,Ghosh:2012sa}
where the attention was centered, mostly, in the change of mass. The obtained
mass shifts split nearly equally
into attractive and repulsive ones, and
range from an attraction of the order of 100-200 MeV
\cite{Tsushima:1998qw,Klingl:1998zj}, through no changes in the mass
\cite{Muhlich:2003tj,Eichstaedt:2007zp}, to a net repulsion of the order of 50
MeV\cite{Lutz:2001mi}. As for the in-medium width of a $\omega$ meson at rest, 
in \cite{Klingl:1997kf}, where the three pion decay mode is
explicitly considered, it is found to
increase by an order of magnitude at normal nuclear
saturation density as compared to the free width, while in a subsequent revision
of the model \cite{Klingl:1998zj} a
value of about 40 MeV is reported. A similar result was also reported in
\cite{Post:2000rf}, while the width was found to be around 60 MeV in
\cite{Riek:2004kx} and
\cite{Muehlich:2006nn}. All these studies point to a considerable increase
of the $\omega$ width in the medium with respect to its free width.

Recently, the use of the hidden gauge theory to deal with the interaction of
vector mesons with baryons, using a unitary approach with coupled channels, has
brought new light into the subject since it interprets some baryonic
resonances as being dynamically generated by this interaction
\cite{sourav,Oset:2009vf}. 
This opens a new framework to study the interaction of vector mesons with
nuclei, as done recently for the $\bar K^*$ \cite{Tolos:2010fq}, where 
a spectacular increase of the width of the $\bar K^*$ in nuclei, of around
250 MeV at nuclear saturation density, was found. With this
perspective, we evaluate in this work the
$\omega$ width in nuclear matter, exploiting
the analogies with the $\bar K^*$ meson and addressing other decay channels
specific to the $\omega$ meson tied to its decay into
three pions.

The paper is organized as follows. 
The formalism employed is described in Sect.~\ref{sec:formalism}, where the
Lagrangians needed for the study of the $\omega$ properties in the medium are
presented. In the next subsections, the expressions for the $\omega$
self-energy in nuclear matter corresponding to the different decay channels
explored in this work are derived. Our results are discussed in
Section \ref{sec:results} and some concluding remarks are given in
Sect.~\ref{sec:conclusions}.

\section{Formalism for vector meson interactions}
\label{sec:formalism}

A free $\omega$ meson decays predominantly into three pions, most of the strength
being associated to the process $\omega \to \rho \pi$ with the subsequent decay of the $\rho$ meson into two pions. Since $m_\omega <
m_\rho+m_\pi$, the mechanism proceeds through the tail of the $\rho$-meson
distribution, as a result of which the $\omega$ width is relatively
small, $\Gamma^{(0)}_\omega=8.49\pm0.08$ MeV, with 89.2\% of this value corresponding to the $3\pi$ decay channel \cite{pdg}. The situation may change drastically in the nuclear medium, since any of
the decay pions can be absorbed by a nucleon, exciting a so-called 
particle-hole excitation and opening the allowed phase-space considerably.
In addition, other mechanisms not forbidden by QCD symmetries but
energetically closed in free space, such as $\omega\to K \bar K$,  may also
contribute when the medium modifications of
the $\bar K$ and $K$ mesons are incorporated. One should also consider the
modification of the $\omega$ properties associated
to its quasielastic collisions with the nucleons in the medium or other inelastic processes not incorporated by the previously mentioned absorption mechanisms.

\begin{figure}[tb]
\includegraphics[width=0.8\textwidth]{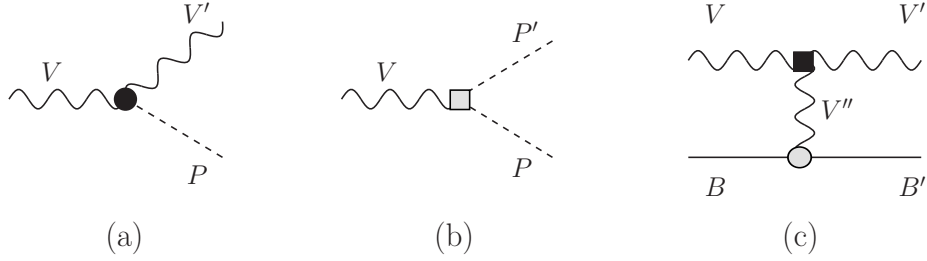}
\caption{Diagrams depicting the vertices needed in the evaluation of the $\omega$ meson self-energy: the anomalous $VVP$ term (a),  the $VPP$ term (b), and the $VB \to V'B'$ amplitude mediated by t-channel vector-meson exchange.}%
\label{fig:vertices}%
\end{figure}

In this section we describe how the different contributions to the $\omega$
self-energy explored in this work are obtained. We start presenting the Lagrangians describing the
interaction of the $\omega$
meson with other mesons (vectors and pseudoscalars) and with baryons of the ground-state octet. 
The transition $\omega \to \rho \pi$, depicted generically in Fig.~\ref{fig:vertices}(a),  is obtained from
the vector meson dominance (VMD) Lagrangians involving
the anomalous $VVP$ term described in
\cite{Bramon:1992kr}
\begin{equation}
{\cal L}_{VVP} = \frac{G}{\sqrt{2}}\epsilon^{\mu \nu \alpha \beta}\langle
\partial_{\mu} V_{\nu} \partial_{\alpha} V_{\beta} P \rangle \ ,
\label{eq:lagranom}
\end{equation}
where $P$ is the matrix of pseudoscalar fields 
\begin{equation}
P =
\left(
\begin{array}{ccc}
\frac{1}{\sqrt{2}} \pi^0 + \frac{1}{\sqrt{6}} \eta_8 & \pi^+ & K^+ \\
\pi^- & - \frac{1}{\sqrt{2}} \pi^0 + \frac{1}{\sqrt{6}} \eta_8 & K^0 \\
K^- & \bar{K}^0 & - \frac{2}{\sqrt{6}} \eta_8
\end{array}
\right) \ ,
\label{eq:phi}
\end{equation} 
$V_\mu$ is the matrix of vector mesons of the nonet of the $\rho$
\begin{equation}
V_\mu=\left(
\begin{array}{ccc}
\frac{\rho^0}{\sqrt{2}}+\frac{\omega}{\sqrt{2}}&\rho^+& K^{*+}\\
\rho^-& -\frac{\rho^0}{\sqrt{2}}+\frac{\omega}{\sqrt{2}}&K^{*0}\\
K^{*-}& \bar{K}^{*0}&\phi\\
\end{array}
\right)_\mu \ ,
\label{Vmu}
\end{equation}
and the $\langle \rangle$ symbol represents the trace in SU(3) space.
The coupling constant of the $VVP$ Lagrangian is $G=\displaystyle\frac{3{g^\prime}^2}{4\pi^2 f}$, where
$g^\prime=-\displaystyle\frac{G_V M_V}{\sqrt{2}f^2}$ \cite{Bramon:1992kr}, with
$f=f_\pi=93\,$MeV and $M_V$ an appropriate vector meson mass, which can be taken equal to the mass of the $\rho$ meson. The value of
 $G_V$ can be adjusted to $\rho \to \pi \pi$ decay ($G_V=69$ MeV), to the pion
charged radius ($G_V=55$ MeV) or can be fixed invoking consistency with QCD
asymptotic behavior ($G_V=f/\sqrt{2}$) \cite{Ecker:1988te,Ecker:1989yg}. In the
latter case one obtains $G=\displaystyle\frac{3{m_\rho}^2}{16\pi^2 f^3}=14$
GeV$^{-1}$. {Alternatively, one can also use the local hidden gauge approach, which in its version of full vector meson dominance also leads to the same result \cite{Harada:2003jx,Benayoun:2009im}.}

The
formalism of the hidden gauge interaction for vector mesons, which we take from 
\cite{hidden1,hidden2} (see also \cite{hidekoroca} for a practical set of
Feynman rules), provides the interaction of 
vector mesons amongst themselves
\begin{equation}
{\cal L}_{III}=-\frac{1}{4}\langle V_{\mu \nu}V^{\mu\nu}\rangle \ ,
\label{lVV}
\end{equation}
 where $V_{\mu\nu}$ is given by 
\begin{equation}
V_{\mu\nu}=\partial_{\mu} V_\nu -\partial_\nu V_\mu -ig[V_\mu,V_\nu]\ ,
\label{Vmunu}
\end{equation}
with the coupling of the theory given by $g=M_V/2f$. 
This Lagrangian gives rise to a three 
vector vertex
\begin{equation}
{\cal L}^{(3V)}_{III}=ig\langle (\partial_\mu V_\nu -\partial_\nu V_\mu) V^\mu V^\nu\rangle
=ig\langle (V^\mu\partial_\nu V_\mu -\partial_\nu V_\mu
V^\mu) V^\nu\rangle
\label{l3Vsimp}\ ,
\end{equation}
needed for the evaluation of the $VB \to V'B'$ interacting terms mediated by
t-channel vector meson exchange, as seen in Fig.~\ref{fig:vertices}(c), which
will describe, after unitarization, the elastic $\omega N \to \omega N$
interaction and the inelastic
transitions to the related coupled channels.
The Lagrangian describing the baryon-baryon-vector vertex involved in the $VB
\to V'B'$
amplitudes is given by
\cite{Klingl:1997kf,Palomar:2002hk} 
\be
{\cal L}_{BBV} =
g \left(\langle\bar{B}\gamma_{\mu}[V^{\mu},B]\rangle+\langle\bar{B}\gamma_{\mu}B\rangle \langle V^{\mu}\rangle \right),
\label{lagr82}
\ee
where $B$ stands for the matrix of the baryon octet \cite{Eck95,Be95}. 
%Similarly,
%one has also a Lagrangian for the coupling of the vector mesons to the baryons
%of the decuplet, which can be found in \cite{manohar}.

Finally, the coupling of the vector to
 pseudoscalar mesons depicted in Fig.~\ref{fig:vertices}(b), needed for
evaluating the $\omega \to \bar K K$ transition, is also provided by the hidden
gauge Lagrangian as
\begin{equation}
{\cal L}_{VPP}= -ig ~\langle [
P,\partial_{\mu}P]V^{\mu}\rangle \ .
\label{lagrVpp}
\end{equation}

\subsection{${\boldmath \omega}$ self-energy from ${\boldmath \rho \pi}$ and uncorrelated ${\boldmath \pi\pi\pi}$}
\label{subsec:rhopi}

\begin{figure}[tb]
\includegraphics[width=0.8\textwidth]{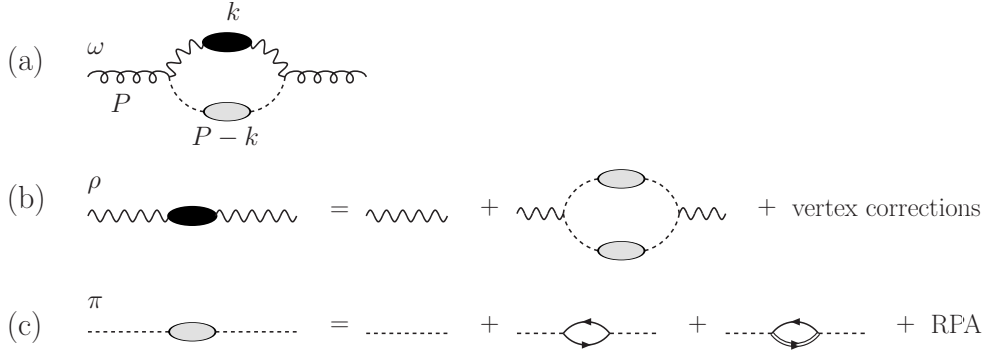}
\caption{Diagrammatic representation of the $\omega$ meson self-energy coming
from its decay into the $\rho\pi$ mode (a), where the $\rho$ meson decays into
two pions (b) and the $\pi$ propagator is dressed by its coupling
to particle-hole and $\Delta$-hole excitations including short-range
correlation effects (c).}%
\label{fig:self_rhopi}%
\end{figure}

Let us consider first the contributions to the $\omega$ width  coming from its
decay into the $\rho\pi$ mode in the nuclear medium. The self-energy for this
process is depicted diagrammatically in Fig.~\ref{fig:self_rhopi}(a), where
the $\rho$- and
$\pi$-meson lines correspond to their medium propagators shown
in Figs.~\ref{fig:self_rhopi}(b) and  (c), respectively.

From the $VVP$ Lagrangian, \ref{eq:lagranom}, we obtain the following
$\omega\rho\pi$ interaction vertex:
\begin{equation}
-iT=i{\cal L}_{\omega\rho\pi} = i G \, \epsilon^{\mu \nu \alpha \beta}\,
\partial_\mu \omega_\nu \partial_\alpha\rho_\beta \,\pi =
 i G \, \epsilon^{\mu \nu \alpha \beta} \, P_\mu \epsilon_\nu k_\alpha\epsilon^\prime_\beta  \ ,
\label{eq:vertexVVP}
\end{equation}
where $P$ and $k$ are the four-momenta of the $\omega$ and $\rho$ mesons,
respectively.
The contribution to the $\omega$ self-energy coming from the diagram of
Fig.~\ref{fig:self_rhopi}(a) can then be written as:
\begin{eqnarray}
-i\Pi_{\omega\to\rho\pi}(\rho,P)=&&\int \frac{d^4 k}{(2\pi)^4} 
i G \, \epsilon^{\mu \nu \alpha \beta} P_\mu \epsilon_\nu k_\alpha\epsilon^\prime_\beta \, 
i G \, \epsilon^{\mu^\prime \nu^\prime \alpha^\prime \beta^\prime} P_{\mu^\prime}
\epsilon_{\nu^\prime} k_{\alpha^\prime} \epsilon^\prime_{\beta^\prime} \nonumber \\
&\times& \,i D_\rho(\rho,k^0,\vec{k}\,) \,
\, i D_\pi(\rho,P^0-k^0,\vec{P}-\vec{k}\,)  \ ,
\label{eq:self_rhopi}
\end{eqnarray}
where $D_\rho$ and $D_\pi$ stand, respectively, for the $\rho$ and $\pi$
propagators. Working out the
index
contractions and using the spectral decomposition of the meson
propagators,
\begin{equation}
D_x(\rho,q^0,\vec{q}\,)=\int_0^\infty \frac{d\omega}{\pi}(-2\omega)\frac{{\rm
Im}\,D_x(\rho,\omega,\vec{q}\,)}{(q^{0})^2-\omega^2+i\varepsilon} ~~~\mbox{with
$x=\rho,\pi$} \ ,
\label{eq:lehmann}
\end{equation}
Eq.~(\ref{eq:self_rhopi}) reduces to
\begin{eqnarray}
\Pi_{\omega\to\rho\pi}(\rho,P) = && \int \frac{d^3 k}{(2\pi)^3}
\int_0^\infty \frac{d\omega}{\pi} \int_0^\infty \frac{d\omega^\prime}{\pi}
\left[\frac{2 G^2\{(k P)^2-P^2 k^2\} }{P^0-\omega-\omega^\prime+i\varepsilon}-
\frac{2 G^2\{(k^\prime P)^2-P^2 k^{\prime\, 2}\}
}{P^0+\omega+\omega^\prime-i\varepsilon}\right]
\nonumber \\
&& \times \, {\rm Im}\, D_\rho(\rho,\omega,\vec{k}\,) \, \, {\rm Im}\,
D_\pi(\rho,\omega^\prime,\vec{P}-\vec{k}\,)  \ ,
\label{eq:self_rhopi2}
\end{eqnarray}
with $P=(P^0,\vec{P}\,)$, $k=(\omega,\vec{k}\,)$, 
$k^{\prime}=(-\omega,\vec{k}\,)$,
after integrating appropriately over $k^0$ and omitting the polarization
dependence $\vec{\epsilon}\,\vec{\epsilon}\,^\prime$.
For practical
purposes, one can take $k^\prime \simeq k$ in the last term of
Eq.~(\ref{eq:self_rhopi2}),
which contributes very little due to the much larger denominator, and then the
expression can be written more compactly as
\begin{eqnarray}
\Pi_{\omega\to\rho\pi}(\rho,P) = && \int \frac{d^3 k}{(2\pi)^3}
\int_0^\infty \frac{d\omega}{\pi} \int_0^\infty \frac{d\omega^\prime}{\pi}
2 G^2\{(k P)^2-P^2 k^2\} 
\frac{2(\omega+\omega^\prime)}{(P^{0})^2-(\omega+\omega^\prime)^2+i\varepsilon}
\nonumber \\
&& \times \, {\rm Im}\, D_\rho(\rho,\omega,\vec{k}\,) \, \, {\rm Im}\,
D_\pi(\rho,\omega^\prime,\vec{P}-\vec{k}\,)  \ .
\label{eq:self_rhopi3}
\end{eqnarray}
Note that {Eqs.~(\ref{eq:self_rhopi2}) and (\ref{eq:self_rhopi3})} contain a factor
of 3 accounting for the three different charged channels, $\rho ^+ \pi^-$ ,
$\rho^0 \pi^0$ and $\rho^- \pi^+$, an aspect that will be discussed later.
The in-medium $\pi$ and $\rho$ propagators are obtained from their corresponding self-energy
through the Dyson equation:
\begin{equation}
D_x(\rho,q^0,\vec{q}\,)=\frac{1}{(q^0)^2-\vec{q}\,^2-m_x^2 -\Pi_x(\rho,q^0,\vec{q}\,)} 
~~~\mbox{with
$x=\rho,\pi$} \ .
\label{eq:prop}
\end{equation}

The pion self-energy, $\Pi_\pi(\rho,q^0,\vec{q}\,)$, is strongly dominated by the $p$-wave coupling to
particle-hole ($ph$) and $\Delta$-hole ($\Delta h$) components, displayed in
Fig.~\ref{fig:self_rhopi}(c), and also contains  a small
repulsive $s$-wave contribution that takes over at small momenta, as
well as contributions from 2$p$-2$h$ excitations, which account for
two-nucleon pion
absorption processes. 
The pion self-energy also accounts for the effect of
repulsive, spin-isospin $NN$ and $N\Delta$ short range
correlations \cite{Oset:1981ih}, which are incorporated through a RPA summed Landau-Migdal type interaction, and
implements the usual monopole form factor at
each $\pi NN$ and $\pi N \Delta$ vertex, 
namely $F_{\pi}(\vec{q}\,^2) = (\Lambda_{\pi}^2 - m_{\pi}^2) /
[\Lambda_{\pi}^2 - q^{0\,2} + \vec{q}\,^2 ]$, with
$\Lambda_{\pi}=1200$~MeV, as needed in the empirical study of $NN$
interactions \cite{Machleidt:1987hj}. The details of the pion
self-energy employed in this work can be found in
Refs.~\cite{Oset:1989ey,Ramos:1994xy}. 

This pion self-energy is employed in Ref.~\cite{Cabrera:2000dx} to obtain, via
the diagrams of Fig.~\ref{fig:self_rhopi}(b), the in-medium $\rho$ self-energy
and the corresponding in-medium propagator. The model also implements vertex
corrections and the effect of resonance-hole excitations, 
finding that, at normal nuclear matter
density $\rho_0=0.17$ fm$^{-3}$, the $\rho$ meson mass does not change and its
width is increased by one third with
respect to its free value, $\Gamma_\rho^{\rm free}=149.4$ MeV. To simplify our
calculations, we parametrize the results of Ref.~\cite{Cabrera:2000dx}
in terms of a simple width function that depends on the nuclear density and is
corrected by the kinematical factors associated to the fact that we are dealing
with an off-shell $\rho$ meson with invariant mass squared
$s=\omega^2-\vec{k}\,^2$. Thus,
\begin{equation}
\Gamma_\rho(\rho,s)=\Gamma_\rho^{(0)}(s)+\Delta\Gamma_\rho(\rho,s) \ ,
\label{eq:rhowidth0}
\end{equation}
where
\begin{equation}
 \Gamma_\rho^{(0)}(s)=\Gamma_\rho^{\rm free}
\frac{m_\rho^2}{s}\frac{\bar{k}^3(s)}{
\bar{k}_{\rm on}^3} \ ,
\label{eq:widthrho0}
\end{equation}
and
\begin{equation}
 \Delta\Gamma_\rho(\rho,s)=0.33\,\Gamma_\rho^{\rm free}
\frac{\rho}{\rho_0}\frac{m_\rho^2}{s}\frac{{
k^\prime}^3(s)} { { k^\prime}_{\rm on}^3}  \ .
\label{eq:rhowidth}
\end{equation}
The momenta appearing in the former equations are given by
\begin{equation}
\bar{k}(s)=\frac{\lambda^{1/2}(s,m_\pi^2,m_\pi^2)}{2\sqrt{s}}\theta(\sqrt{s}
-2m_\pi) \ ,~~~~ \bar{k}_{\rm on}= \bar{k}(m_\rho^2)\ , \nonumber
\end{equation}
\begin{equation}
{k^\prime}(s)=\frac{\lambda^{1/2}(s,m_\pi^2,0)}{2\sqrt{s}}\theta(\sqrt{s}-m_\pi)
\ ,~~~~  {k^\prime}_{\rm on}={k^\prime}(m_\rho^2) \ , \nonumber
\end{equation}
where $\lambda(x,y,z)=x^2+y^2+z^2-2xy-2yz-2zx$ is the K\"allen function. Note
that the expression for $k^\prime$ reflects the fact
that the in-medium $\rho$ meson decays, in first approximation, into a $\pi$
meson and a
$ph$ excitation of essentially zero energy. The $\rho$ meson self-energy needed
for the evaluation of the in-medium $\rho$ propagator in
Eq.~(\ref{eq:self_rhopi3}) is
then given by
\begin{equation}
\Pi_\rho(\rho,k)=-i\, m_\rho\, \Gamma_\rho(\rho,\omega^2-\vec{k}\,^2) \ . 
\end{equation}

This approach of parametrizing the $\rho$ spectral function in terms of simply increasing the $\rho$ width has been employed in some experimental analyses. However, it misses the bump in the $\rho$ spectral function at low invariant masses due to the $N^*(1520) h$ excitation \cite{Peters:1997va,Cabrera:2000dx}, a  region of the $\rho$ spectral function that plays a relevant role in the $\omega$ self-energy as shown recently in \cite{Cabrera:2013zga}. On the other hand, recent developments on the vector-baryon interaction using unitarity in coupled channels and input from the local hidden gauge approach \cite{Oset:2009vf, Garzon:2012np,Garzon:2013pad} are able to provide the $\rho N$ scattering amplitude, while generating dynamically some resonances at the same time, such as the $N^*(1520)$ and the $N^*(1700)$ with $J^P=3/2^-$ \cite{Garzon:2013pad} and one related to the $N^*(1650)$ with $J^P=1/2^-$ \cite{Oset:2009vf, Garzon:2012np}. The isospin $I=3/2$ amplitudes are also evaluated in \cite{Oset:2009vf}, although they do not present resonant structures (some resonances are generated from the interaction of vector mesons with the decuplet of baryons \cite{sourav} but they do not apply here).

In view of this, we will also perform calculations dressing the $\rho$ with its low-density self-energy
\begin{eqnarray}
\Pi_{\rho} = t_{\rho N\to \rho N} \rho \ ,
\end{eqnarray}
where $t_{\rho N\to \rho N} $ is the $\rho N$ amplitude calculated in \cite{Oset:2009vf, Garzon:2012np,Garzon:2013pad} averaged over total isospin $I$ and angular momentum $J$, 
\begin{eqnarray}
t_{\rho N\to \rho N} = \frac{\displaystyle\sum_{I,J} (2I+1) (2J+1)t_{\rho N\to \rho N} ^{I\,J}} {\displaystyle\sum_{I,J} (2I+1)(2J+1)} \ , 
\end{eqnarray}
with $I=1/2,3/2$ and $J=1/2,3/2$. 

\begin{figure}[t]
\begin{center}
\includegraphics[width=0.6\textwidth]{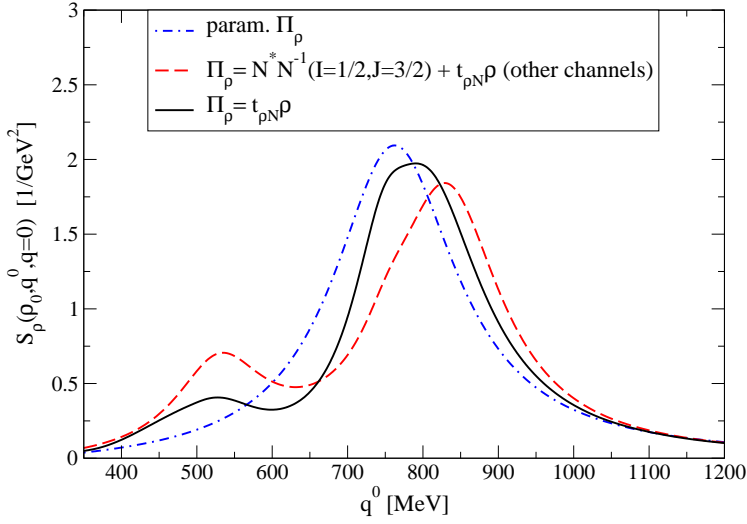}
\caption{The spectral function of a $\rho$-meson of zero momentum in nuclear
matter at saturation density for the three prescriptions employed in this paper.
Dash-dotted line: the phenomenological width given by
Eqs.~(\ref{eq:rhowidth0})-(\ref{eq:rhowidth});  dashed-line: employing the
$t_{\rho N\to \rho N}$ model from the coupled channel unitary model with local
hidden gauge lagrangians of Refs.~\cite{Oset:2009vf,
Garzon:2012np,Garzon:2013pad} but replacing the $I=1/2,J^P=3/2^-$ amplitude by
the $N^*(1520)N^{-1}$ contribution of Ref.~\cite{Cabrera:2000dx}; solid line:
taking the complete  $t_{\rho N\to \rho N}$ amplitude from the model of
Refs.~\cite{Oset:2009vf, Garzon:2012np,Garzon:2013pad}.
}
\label{fig:rho_prop}
\end{center}
\end{figure}

In Fig.~\ref{fig:rho_prop} we show results for the $\rho$ spectral function obtained from three prescriptions for the $\rho$ self-energy:\\
(a) dash-dotted line: the phenomenological width given by Eqs.~(\ref{eq:rhowidth0})-(\ref{eq:rhowidth}) , \\
(b)  dashed line: employing the amplitude $t_{\rho N\to \rho N}$ obtained in Refs.~\cite{Oset:2009vf, Garzon:2012np,Garzon:2013pad} from a coupled channel unitary approach with input from the local hidden gauge lagrangians,  but replacing the $I=1/2,J^P=3/2^-$ contribution by the $N^*(1520)N^{-1}$ term of Ref.~\cite{Cabrera:2000dx} that is parametrized as
\begin{eqnarray}
t^{1/2,3/2^-}=\frac{\tilde{g}_{N^* N \rho}^2}{\sqrt{s^{\prime}}-M_R +{\rm i} \Gamma/2} \ ,
\end{eqnarray}
where $M_R$=1520 MeV, $\Gamma=150$ MeV and $\tilde{g}_{N^*N \rho}=7.73$, a value rather similar to the one obtained in \cite{Garzon:2013pad}, $|\tilde{g}_{N^*N \rho} | =6.39$. The argument $\sqrt{s^{\prime}}$ of $t^{1/2,3/2^-}$ is determined from the four-momentum of the $\rho$, $p_{\rho}=(\omega,\vec{k}\,)$, and that of a nucleon at rest in the nucleus,
$s^{\prime}=(p_{\rho}+p_{N})^2=(\omega+m_N)^2-\vec{k}\,^2$, and\\
(c) solid line: taking the complete $t_{\rho N\to \rho N}$ from the model of Refs.~\cite{Oset:2009vf, Garzon:2012np,Garzon:2013pad}

We can see in Fig.~\ref{fig:rho_prop} that the $\rho$ width is quite similar in all cases, but the bump from the $N^*(1520)$ excitation is missing in the pure phenomenological model based on Eqs.~(\ref{eq:rhowidth0})-(\ref{eq:rhowidth}). This excitation has a larger strength in the $N^* N^{-1}$ resonance pole approach of  \cite{Cabrera:2000dx} than in the microscopical model of \cite{Garzon:2013pad}, a difference that will have repercussions on the value of the $\omega$ width in the medium as will see.

Finally, the in-medium $\omega$ width coming from the $\omega \to \rho \pi$ channel is  obtained from
\begin{equation}
\Gamma_{\omega\to\rho\pi}(\rho,P)=-\frac{{\rm Im}\,
\Pi_{\omega\to\rho\pi}(\rho,P)}{P^0}  \ ,
\end{equation}
and the corresponding in-medium width correction is
\begin{equation}
\Delta\Gamma_{\omega\to\rho\pi}(\rho,P)= \Gamma_{\omega\to\rho\pi}(\rho,P) -
\Gamma^{(0)}_{\omega\to\rho\pi}(P) \ ,
\end{equation}
where the free $\omega \to \rho \pi$ decay width can be obtained from the $\omega$
self-energy of Eq.~(\ref{eq:self_rhopi3}) by
replacing
\begin{equation}
{\rm Im}\, D_{\pi}(\rho,\omega^\prime,\vec{P}-\vec{k}\,) \to
-\pi\delta(\omega^\prime-\omega_\pi(\vec{P}-\vec{k}\,)) \frac{1}{2
\omega_\pi(\vec{P}-\vec{k}\,)} \ ,
\label{eq:freepi}
\end{equation}
\begin{equation}
{\rm Im}\, D_\rho(\rho,\omega,\vec{k}\,) \to
{\rm Im}\,\frac{1}{\omega^2-\vec{k}^2-m_\rho^2+ i
m_\rho\Gamma_\rho^{(0)}(\omega^2-\vec{k}^2)} \ .
\label{eq:freerho}
\end{equation}
with $\omega_\pi(\vec{P}-\vec{k}\,)=\sqrt{(\vec{P}-\vec{k}\,)^2+m_\pi^2}$.
After performing the integration over $\vec{k}$, an appropriate change of variable allows one to write the free $\omega \to \rho\pi$ width as:
\begin{equation}
 \Gamma^{(0)}_{\omega\to\rho\pi}(P)=-\frac{{\rm
Im}\,\Pi_{\omega\to\rho\pi}^{(0)}(P)}{P^0} =
-\frac{G^2}{4\pi^2} \int d\tilde{m}_\rho^2 \, {\rm
Im}\,\frac{k^3(\tilde{m}_\rho^2)}{\tilde{m}_\rho^2-m_\rho^2 + i m_\rho
\Gamma_\rho^{(0)}(\tilde{m}_\rho^2) }\ ,
\label{eq:widthfree}
\end{equation}
where $k(\tilde{m}_\rho^2)=\lambda^{1/2}(m^2_\omega, \tilde{m}^2_\rho,
m_\pi^2)\theta(m_\omega-\tilde{m}_\rho-m_\pi)/(2 m_\omega)$.

Note that the factor of 3 in  Eqs.~(\ref{eq:self_rhopi2})-(\ref{eq:self_rhopi3}) would account appropriately for the three different isospin channels, $\rho ^+ \pi^-$ ,
$\rho^0 \pi^0$ and $\rho^- \pi^+$, if
the $\rho$
meson was a stable particle, since in this case the
three channels would correspond to different final states that would not interfere.
\begin{figure}[htb]
\includegraphics[width=0.8\textwidth]{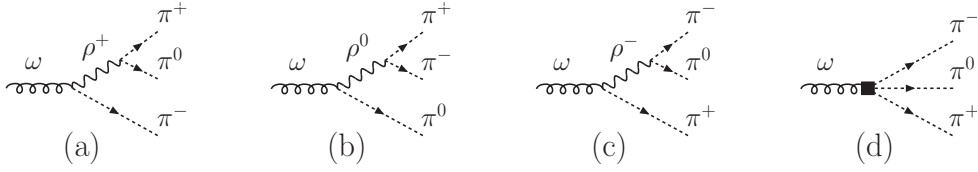}
\caption{Contributions to the  $\omega \to \rho \pi \to \pi^+ \pi^0 \pi^+$ decay
process, coming from three different intermediate charged channels:
$\rho^+ \pi^-$, $\rho^0 \pi^0$  and $\rho^- \pi^+$.} %
\label{fig:vert_rhopi}%
\end{figure}
However, the final $\rho$ meson decays into two pions, as depicted by the first three diagrams of
Fig.~\ref{fig:vert_rhopi}. 
\begin{figure}[hbt]
\includegraphics[width=0.8\textwidth]{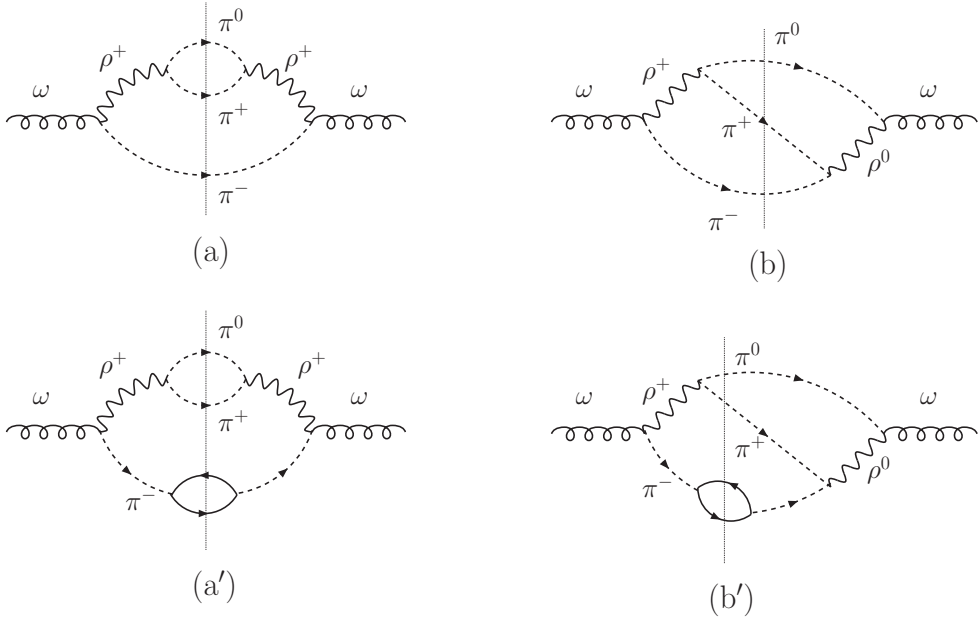}
\caption{Diagrammatic representation of direct and interference
contributions to the $\omega$ meson self-energy coming from its decay into the
$\rho\pi$ mode. Top row: free space contributions. Bottom row: in-medium
contributions.}%
\label{fig:interf}%
\end{figure}
One then has
identical final states coming from the three different channels that may
give rise to important interference effects. This is more clearly visualized from
the contributions to the $\omega$ width in free space displayed in the
top row of
Fig.~\ref{fig:interf}, where the direct contribution of the $\rho ^+ \pi^-$
channel is shown in diagram (a), while the interference between the
$\rho ^+ \pi^-$ and $\rho^0 \pi^0$ channels is shown in diagram (b).
There are a total of 9 contributions of the type shown in the top row
of Fig.~\ref{fig:interf}, but in our calculation we are only considering the
three direct terms of type (a)
and, consequently, we will miss part of the
free $\omega$ width, as we will see. Since the interference terms vanish when the $\rho$ width is zero, a rough guess of their relevance can be given by $\Gamma_{\rho}/m_{\rho}$. However, when these processes occur in the
medium,
the situation is quite different, as can be easily understood from the diagrams
displayed in the bottom row of Fig.~\ref{fig:interf}. We see there that
the pion emitted in the $\omega$ decay converts into a $ph$ excitation, which
costs very little energy. In this case, the virtual $\rho$ meson can be left
practically on-shell, enhancing
the size of its propagator. For this reason, the direct diagram
represented in Fig.~\ref{fig:interf}(a$^\prime$), where the two $\rho$ propagators can be practically
put on-shell simultaneously, is much larger than any of the other interference
diagrams, such as that in Fig.~\ref{fig:interf}(b$^\prime$),
because in this case when the $\rho$ propagator on the left carries practically the full incoming energy $m_\omega$, being then almost on-shell, the $\rho$
propagator on the right only
carries an energy of about $m_\omega/2$, being far off-shell. Therefore, the contribution of
the interference terms to the in-medium corrections of the $\omega$ meson
self-energy is essentially negligible. A more quantitative estimate can be done
knowing that the average $\pi^-$ momentum in diagrams (a$^\prime$) and
(b$^\prime$) carries around 300 MeV/c. One finds that the $ph$ excitation takes
about 50 MeV of energy and then the $\rho^+$ has an invariant mass of about 665
MeV in diagrams (a$^\prime$) and (b$^\prime$), while the $\rho^0$ in diagram
(b$^\prime$) has an invariant mass around 387 MeV. The combination of the
reduced $\rho^0$ propagator and its smaller $\pi \pi$ vertex renders the diagram
(b$^\prime$) much less relevant compared to (a$^\prime$) than diagram (b) is
with respect to (a), since in this latter case the invariant masses of the
propagators are much closer, as one finds, analogously, taking an
estimated average $\pi^-$ momentum of 200 MeV/c.

In order to obtain the $\omega$ width, we need to take a value for the coupling
constant $G$. Following the arguments given in the recent work of
Ref.~\cite{Gudino:2011ri}, this coupling can be determined from different
sources. Their fit to vector meson radiative decays considering that the photon
emission is mediated by neutral vector mesons favors a value $G=11.9\pm 0.2$
GeV$^{-1}$, which is supported by the agreement between $VMD$ and low
energy theorems for $\pi^0 \to \gamma \gamma$ decay. 
Using this value for the coupling constant $G$, a free $\omega$ width of
$\Gamma_{\omega\to \rho\pi}^{(0)}=4.4$ MeV is reported in \cite{Gudino:2011ri}.
In
the present work we obtain, from equation Eq.~(\ref{eq:widthfree}), a value
of 2.8 MeV, the difference coming precisely from the interference terms not considered here. The
lack of agreement with the experimental $\omega$ width from its three-pion decay
channel,
$\Gamma_{\omega\to 3\pi}^{{\rm exp}}=7.57\pm0.07$ MeV \cite{pdg}, drove the
authors of \cite{Gudino:2011ri} to complement their model with a contact term,
as that of the diagram of Fig.~\ref{fig:vert_rhopi}(d), or 
with contributions from higher mass meson resonances, like the
$\rho^\prime(1450)$. The latter contributions can also be mimicked by a
contact
term, especially because some of the $\rho^\prime$ meson decay parameters are
not sufficiently well known. The coupling constant of the contact term can then
be adjusted so as to reproduce the free decay width of the $\omega$ into three
pions. One may view the effect of the contact term as providing a background 
that needs to be added to the $\omega \to \rho \pi$ process. A different approach to this problem is followed in \cite{Harada:2003jx,Benayoun:2009im}. In these works the local hidden gauge approach is used to determine the anomalous sector and, in addition to the $\omega \rightarrow \rho \pi \rightarrow \pi \pi \pi$ decay mode, one derives a contact term given in terms of the parameters $c_1$, $c_2$ and $c_3$, which one can determine under the assumption of vector meson dominance. Here, we have chosen to follow the more phenomenological approach of \cite{Gudino:2011ri}.

An straightforward implementation of the contact term is obtained by the following replacement
\begin{equation}
G^2 \, {\rm Im}\, D_\rho(\rho,\omega,\vec{k}\,) \to  A
\, {\rm Im}\left\{ \frac{1}{\omega^2-\vec{k}\,^2-m_\rho^2-m_X^2 - i m_\rho
\Gamma_\rho(\rho,s=\omega^2-\vec{k}\,^2)} \right\} \ ,
\label{eq:contact}
\end{equation}
in Eq.~(\ref{eq:self_rhopi3}),
taking $m_X \to \infty$. This substitution implements a
propagator having a practically constant real part and an imaginary part that
incorporates the correct two-pion phase-space in the free case or the medium
corrected phase-space at finite density.

The dynamical arguments
for the negligible effect of in-medium interferences in the case of the $\omega
\to \rho \pi $ process, based on the $\rho$  propagator being practically
on-shell, are no longer valid for the terms involving the contact term (direct contact term contributions and their interference with the $\rho \pi$ terms). We
then assume these contact-term interferences
to have a similar size in free
space and in the medium. Therefore, interpreting the constant $A$ as
containing effectively the effect of these interferences, we adjust its value so that our
calculation reproduces
the strength of the free $\omega$ width not accounted for by the $\omega \to
\rho\pi$ terms, namely  $\Delta
\Gamma^{\rm contact}_{\omega \to 3\pi}= \Gamma_{\omega\to
3\pi}^{{\rm exp}} - \Gamma_{\omega\to \rho\pi}^{(0)} = 7.57$ MeV $- 4.4$
MeV$ = 3.17$ MeV, where we have taken the proper value for $\Gamma_{\omega\to
\rho\pi}^{(0)}$ that contains the interference effects of the $\rho\pi$ mechanism \cite{Gudino:2011ri}.
An alternative option is to adjust the size of $G$ to reproduce the
complete free $\omega \to \pi\pi\pi$ width directly from the $\rho \pi$
mechanism. The
required value is then $G= 15.7$ GeV$^{-1}$  \cite{Gudino:2011ri}.  We shall exploit both options in the results section.

\subsection{${\boldmath \omega}$ self-energy from ${\boldmath \bar{K}K}$}

\begin{figure}[t]
\begin{center}
\includegraphics[width=0.8\textwidth]{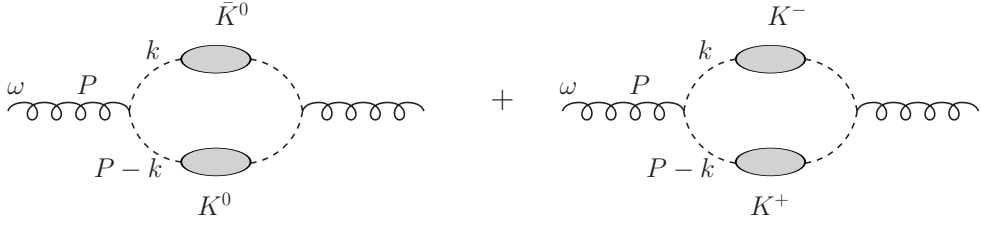}
\caption{The $\omega\to \bar{K}K$ channel renormalized in the nuclear medium. }
\label{fig:omegaloop}
\end{center}
\end{figure}

Analogously to the study of the $\bar{K}^*$ meson in the nuclear medium
\cite{Tolos:2010fq}, where the decay of the $\bar K^*$ meson to $\bar{K}\pi$ is
renormalized by the inclusion of the $\bar{K}$ and $\pi$ self-energies in the
loop function, we consider in this section the
possible modifications of the $K\bar{K}$ loop in the $\omega$ propagator in the
nuclear medium as depicted in Fig. \ref{fig:omegaloop}.   The difference 
respect to the case of the $\bar{K}^*$ meson decaying into $\bar{K}\pi$ is,
though, that the $\omega\to \bar{K} K$ channel is energetically closed in free
space. However, due to the inclusion of the self-energy of the $\bar K$ meson,
some processes are possible in nuclear matter. 

Let us consider first the diagrams in Fig. \ref{fig:omegaloop} in the free case, i.e. ignoring the self-energy insertions.
For the coupling of the $\omega$ meson to a $\bar{K}K$ pair, we take the
Lagrangian given by Eq.~(\ref{lagrVpp}).
Both channels depicted in Fig. \ref{fig:omegaloop}, neutral and charged, are
included in the evaluation of this contribution to the $\omega$ self-energy
that in the free space can be written as
\begin{eqnarray}
-i\Pi_{\omega\to K\bar{K}}^{0}(P)= 4g^2\int
\frac{d^4 k}{(2\pi)^4}  \frac{k_\mu \eps^\mu k_\nu \eps^\nu}{k^2-m_{\bar
K}^2+i\eps}\frac{1}{(P-k)^2-m_{K}^2+i\epsilon}\ .
\label{eq:freeint}
\end{eqnarray}
Making the substitution:
\begin{eqnarray}
k_\mu k_\nu \eps^\mu \eps^\nu \to \frac{1}{3P^2}\left\{ P^2 k^2-(k
P)^2\right\}\, \eps\,\eps^\prime \to\frac{1}{3P^2}\left\{ (k P)^2-P^2
k^2\right\} \vec{\epsilon}\,\vec{\epsilon}\,^\prime \ ,
\end{eqnarray}
which for a $\omega$ meson of zero momentum equals $\vec{k}\,^2/3$, and
performing the integral in the $k^0$ variable, we obtain,
\begin{eqnarray}
\Pi_{\omega\to K\bar{K}}^{0}(P)=\frac{4g^2}{3P^2}\int
\frac{d^3k}{(2\pi)^3}
\left[\frac{\omega_{K}(\vec{P}-\vec{k}\,)+\omega_{\bar{K}}(\vec{k}\,)} {2
\omega_{K}(\vec{P}-\vec{k}\,)\omega_{\bar{K}}(\vec{k}\,)}\right] 
\frac{ (k P)^2-P^2k^2}{(P^0)^2 -
\left[\omega_K(\vec{P}-\vec{k}\,)+\omega_{\bar{K}}(\vec {k }
\,)\right]^2+i\varepsilon}\
.\label{eq:freeself}
\end{eqnarray}
In the medium, we include the self-energy of the antikaon, $\Pi_{\bar K}
(\rho,k^0,\vec{k}\,)$, and obtain the antikaon propagator
\begin{eqnarray}
D_{\bar K}(\rho,k^0,\vec{k}\,)=\frac{1}{(k^0)^2-\vec{k}^2-m_{\bar K}^2-\Pi_{\bar K}
(k^0,\vec{k}\,)} \ ,
\end{eqnarray}
that in the Lehmann representation reads
\begin{eqnarray}
D_{\bar{K}}(\rho,k^0,\vec{k}\,)=\int^\infty_0\frac{d\omega}
{\pi}(-)
\left\{\frac{\mathrm{Im}
D_{\bar{K}}(\rho,\omega,\vec{k}\,)}{k^0-\omega+i\eta}-\frac{\mathrm{Im}
D_K(\rho,\omega,\vec{k}\,)}{k^0+\omega-i\eta}\right\}\ .\label{eq:antikprop}
\end{eqnarray}
For the details of the antikaon self-energy and, 
hence, its propagator,
we refer to
\cite{angels,bennhold,inoue,ollerulf,Ramos:1999ku,Tolos:2006ny,Cabrera:2009qr,
Oset:1989ey,Ramos:1994xy,Waas:1996fy,Waas:1997pe,GarciaRecio:2002cu}. 
%[65-74].
Essentially, the $\bar{K}$ self-energy in symmetric nuclear matter is obtained
from the antikaon-nucleon interaction within a chiral unitary approach. The
model incorporates $s$- and $p$-wave contributions in a self-consistent 
manner. The coupled-channel structure for the $s$-wave interaction includes the
following channels: $\bar{K}N$, $\pi\Sigma$, $\eta\Lambda$, $K\Xi$ for isospin $I=0$,
and $\bar{K}N$, $\pi\Lambda$, $\pi\Sigma$, $\eta\Sigma$, $K\Xi$ for isospin $I=1$. The
$p$-wave contribution to the $\bar{K}$ self-energy is built up from the coupling
of the $\bar{K}$ meson to $\Lambda N^{-1}$, $\Sigma N^{-1}$ and $\Sigma^*
N^{-1}$ excitations.

For the kaon, due to the much weaker $KN$ interaction, we use the low-density approximation and replace $m_K^2$ $\to$
$m_K^2+t_{KN\to KN}\,\rho$, with $t_{KN\to KN}\,\rho=0.13\, m_K^2\rho/\rho_0$
\cite{Oset:2000eg}. Thus, we can write the self-energy of the
$\omega$ meson as,
\begin{eqnarray}
-i\Pi_{\omega\to K\bar{K}}(\rho,P)=&&- \frac{4g^2}{3P^2}\int \frac{d^4
k}{(2\pi)^4} \frac{
(k P)^2-P^2 k^2}{(P-k)^2-m^2_K-t_{KN\to KN}\,\rho}\nonumber\\
&& \times \, \int^\infty_0\frac{d\omega}{\pi} \left\{\frac{\mathrm{Im}
D_{\bar{K}}(\omega,\vec{k}\,)}{k^0-\omega+i\eta}-\frac{\mathrm{Im}
D_K(\omega,\vec{k}\,)}{k^0+\omega-i\eta}\right\} \ .
\end{eqnarray}
As we take the physical mass of the $\omega$ meson, the real part of its
in-medium self-energy must vanish at $\rho=0$.
%The part of negative energy of the $\bar{K}$ propagator, the part of the
%$K$, with $\pi$ in the numerator, is small, its medium modifications are not
%relevant and to a very good %approximation we can cancel it with the
%subtraction
%of the free term. In fact,
%\begin{eqnarray}
%& &\frac{1}{\pi^2}\int^\infty_0 d\omega \,\mathrm{Im}
%D_\pi(\omega,\vec{q}\,)\frac{\pi}{2\omega_K(P-q)(P^0+\omega+\omega_K(P-q))}
%\nonumber\\&\sim&
%\frac{1}{\pi^2}\int^\infty_0
%d\omega\frac{1}{2\omega_\pi(q)}(-)\pi\,\delta(\omega-\omega_\pi(q))\frac{\pi}{
%2\omega_K(P-q)(P^0+\omega+\omega_K(P-q))}\nonumber\\&\equiv&-
%\frac{1}{2\omega_\pi(q)}\frac{1}{2\omega_K(P-q)}\frac{1}{
%P^0+\omega_\pi(q)+\omega_K(P-q)}\ %,
%\end{eqnarray}
%which is the same result as the second term in Eq. (\ref{eq:free}). Since we
%want to get %the $\bar{K}^*$ mass at its physical value at $\rho=0$
Therefore, 
the real part of the free $\omega$ self-energy, $\Pi^{(0)}_{\omega\to K\bar{K}}(P)$,
must be subtracted to $\Pi_{\omega\to K\bar{K}}(\rho,P)$. 
The part of negative energy of the $\bar{K}$ propagator, represented by the last
term in Eq.~(\ref{eq:antikprop}), is small. 
Its medium modifications are not
relevant and to a very good approximation it will be cancelled by
the corresponding term of the free $\omega$ self-energy.
Then, after integrating over the $k^0$ variable, and subtracting
the real part of the free $\omega \to K \bar{K}$ self-energy of Eq.~(\ref{eq:freeself}), one gets the following correction to the $\omega$ width:
\begin{eqnarray}
\Delta\Pi_{\omega\to K\bar{K}}(\rho,P)=&&-\frac{4g^2}{3P^2}\int
\frac{d^3k}{(2\pi)^3} \left\{ (k P)^2-P^2
k^2 \right\} \left[ \int^\infty_0 \frac{d\omega}{\pi}\frac{\mathrm{Im}
D_{\bar{K}}(\rho,\omega,k)}{2\tilde{\omega}_K(\vec{P}-\vec{k}\,)}\frac{1}{
P^0-\tilde{\omega}_K(\vec{P}-\vec{k}\,)-\omega+i\eta}\right.\nonumber\\
~~~~~~&&\left.+\frac{1}{2\omega_K(\vec{P}
-\vec{k}\,) 2\omega_{\bar
K}(\vec{k}\,)}\, \frac{1}{P^0-\omega_K(\vec{P}-\vec{k}\,)-\omega_{\bar
K}(\vec{k}\,)+i\eta}\right] \ .
\label{eq:kaon}
\end{eqnarray}
where $\tilde{\omega}_K(\vec{P}-\vec{k})=\sqrt{(\vec{P}-\vec{k})^2+m_{K}^2 +
t_{KN\to KN}\,\rho}$.
\begin{figure}[ht]
\begin{center}
\includegraphics[width=0.8\textwidth]{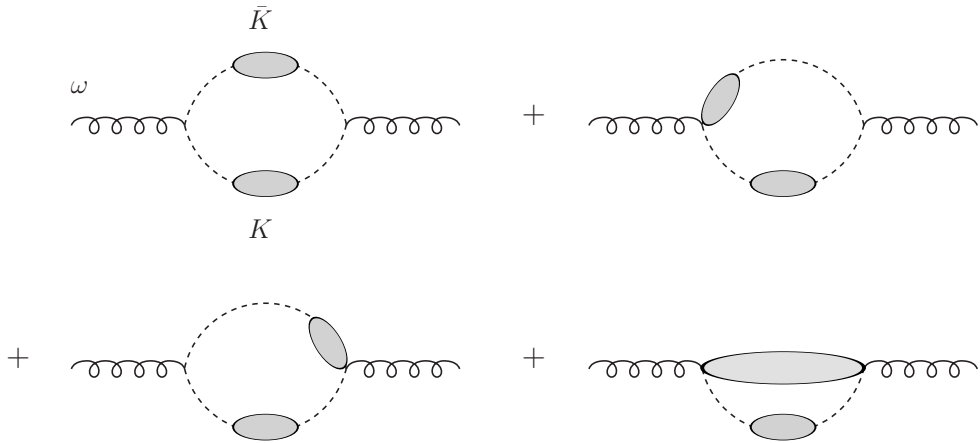}
\caption{Self-energy diagrams at first order in the nuclear density contributing
to the decay of the $\omega$ meson in the medium. }
\label{fig:3}
\end{center}
\end{figure}
Because of gauge invariance of the model, it is necessary to include vertex
corrections \cite{Herrmann:1993za,Chanfray:1993ue} associated to the
last three diagrams of Fig.~\ref{fig:3}. Here we only consider vertex
corrections for the antikaon. As described in Ref.~\cite{Oset:2000eg}, the vertex corrections are easily implemented by
evaluating the diagram of Fig.~\ref{fig:3} with the
$p$-wave antikaon self-energy replaced by
\begin{eqnarray}
\Pi_{\bar{K}}^{\rm p-wave}(\rho,k) \rightarrow
\frac{\Pi_{\bar{K}}^{\rm p-wave}(\rho,k)}{\vec{k}\,^2} \left(
\vec{k}\,^2+\left[D_{\bar{K}}^{0}(k)\right]^{-1}+
\frac{3}{4}\frac{\left[D_{\bar{K}}^{0}(k)\right]^{-2}}{\vec{k}\,^2}
\right)\ ,
\end{eqnarray}
%with $\left[D_{\bar{K}}^{0}(q)\right]^{-1}=(q^0)^2-\vec{q}\,^2-m^2_K$
where $D_{\bar{K}}^{0}(k)$ is the free $\bar{K}$ meson propagator.

\subsection{${\boldmath \omega}$ self-energy from the s-wave $\omega N$
interaction with vector mesons and baryons}

For completeness we also explore the contribution to the $\omega$ self-energy
coming from the s-wave $\omega N$ interaction with vector mesons and baryons.
The $\omega N$ interaction can be constructed within the hidden gauge formalism
in coupled channels, as done in Ref.~\cite{Oset:2009vf}. We proceed as in this
latter reference by constructing the Feynman diagrams that lead to the
vector-baryon ($VB$) transitions $VB \to V^\prime B^\prime$ via the exchange of
a vector meson, as seen in  Fig.~\ref{fig:vertices}(c). We thus make use
of the three-vector vertex Lagrangian,  given by Eq.~(\ref{l3Vsimp}), as well as the Lagrangian for the coupling of vector mesons to the baryon octet given in
Eq.~(\ref{lagr82}). Since we are interested in studying the interaction of the
$\omega$ meson in
nuclear matter, we concentrate in the isospin $I=1/2$ sector, where we find five
vector
meson-baryon channels that couple to the $\omega N$ system: $\rho N$, $\omega
N$, $\phi N$, $K^*\Lambda$ and $K^* \Sigma$. 

As discussed in Ref.~\cite{Oset:2009vf}, one can proceed
by neglecting the three momentum of the external vectors versus the vector
mass, in a similar way as done for chiral Lagrangians under the low-momentum
approximation, and one obtains the $VB \to V^\prime B^\prime$ transition potential: 
\begin{equation}
V_{i j}= - C_{i j} \, \frac{1}{4 f^2} \, \left( k_i^0 + k_j^0\right)
~\vec{\epsilon}_i\,\vec{\epsilon }_j, \label{kernel} 
\end{equation} 
where
$k_i^0, k_j^0$ are the energies of the incoming and outgoing vector
mesons, respectively, $\vec{\epsilon}_i\,\vec{\epsilon}_j$ is the product
of their
polarization vectors, and $C_{ij}$ are the symmetry coefficients
\cite{Oset:2009vf}.

The meson-baryon scattering amplitude is then obtained from the coupled-channel
on-shell Bethe-Salpeter equation \cite{angels,ollerulf}
\begin{equation} 
T = [I - V \, G_{VB}]^{-1}\, V \ ,
\label{eq:Bethe}
\end{equation}
with $G_{VB}$ being the diagonal matrix of the loop
functions for the meson-baryon intermediate states. We note that the iteration of diagrams implicit in the Bethe-Salpeter
equation
for the scattering of vector mesons with baryons implies a sum over the polarizations of the
internal vector mesons which, because they are tied to the external ones
through the  $\vec{\epsilon}_i\,\vec{\epsilon}_j$  factor, leads to a
correction in the $G_{VB}$ function of $\vec{q}\,^{2}/3 M_V^2$ \cite{luisaxial}, which can be safely neglected in consonance with the low-momentum
approximation
done in \cite{Oset:2009vf}. This leads to the factorization  of  the factor
$\vec{\epsilon}_i\,\vec{\epsilon}_j$ for the
external vector mesons also in the $T$ matrix. As a consequence,  the
$J^P=1/2^-,3/2^-$ vector-baryon scattering amplitudes are degenerate.
%This method provides degenerate pairs of
%resonances which have  $J^P=1/2^-,3/2^-$, a pattern that seems to be reproduced
%by
%the existing experimental data.

The loop function of a vector meson of mass $m$ and a baryon of mass $M$
with total
four-momentum $P$ ($s=P^2$) reads:
\begin{eqnarray} G_{VB}(s,m^2,M^2) &=& i\, 2 M
\int \frac{d^4 q}{(2 \pi)^4} \, \frac{1}{(P-q)^2 - M^2 + {\rm i} \epsilon} \,
\frac{1}{q^2 - m^2 + {\rm i} \epsilon}   \ ,
\label{eq:gpropdr}
\end{eqnarray}
which is evaluated using dimensional regularization and taking a natural value of $-2$ for the subtraction constants
at a regularization scale $\mu=630$ MeV \cite{ollerulf}. However, the
relatively large decay width of the $\rho$  vector meson into $\pi\pi$ pairs
is incorporated in the loop function via the
convolution \cite{nagahiro}:
\begin{eqnarray}
\tilde{G}(s)= \frac{1}{N}\int^{(m+\Delta_r)^2}_{(m-\Delta_l)^2}d\tilde{m}^2
\left(-\frac{1}{\pi}\right)
{\rm Im}\,\frac{1}{\tilde{m}^2-m^2+{\rm i} m \Gamma^{(0)}(\tilde{m}^2)}
& G_{VB}(s,\tilde{m}^2,M^2)\ ,
\label{Gconvolution}
\end{eqnarray}
with
\begin{equation}
N=\int^{(m+\Delta_r)^2}_{(m-\Delta_l)^2}d\tilde{m}^2
\left(-\frac{1}{\pi}\right){\rm Im}\,\frac{1}{\tilde{m}^2-m^2+{\rm i}m
\Gamma^{(0)}(\tilde{m}^2)}
\label{Norm}
\end{equation}
being the normalization factor. The integration range around the $\rho$ mass is
established by the left and right parameters $\Delta_l$, $\Delta_r$, 
taken to be a few
times the free $\rho$ width, $\Gamma^{\rm free}_\rho$. The energy dependent width
$\Gamma^{(0)}(\tilde{m}^2)$ is given in Eq.~(\ref{eq:widthrho0}).

\begin{figure}[ht]
\begin{center}
\includegraphics[width=0.35\textwidth]{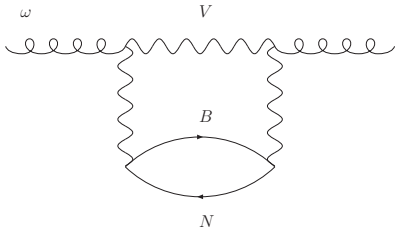}
\caption{Self-energy diagram coming from the s-wave $\omega N$ interaction with
vector mesons and baryons}
\label{fig:diag}
\end{center}
\end{figure}

In nuclear matter we should consider the contributions to the $\omega$
self-energy
coming from its interactions with the nucleons in the Fermi sea. The weak
interaction of $\omega$ mesons with the nucleons, as seen in
Ref.~\cite{Oset:2009vf}, suggests the possibililty of extending the validity of
the low-density theorem to normal nuclear matter density or beyond. Thus,  we
obtain the $\omega$ self-energy, depicted in Fig.~\ref{fig:diag},
by summing the ${t}_{\omega N \to \omega N}$ amplitude in free space over the
nucleon Fermi sea,
$n(\vec{k}\,)=\theta(k_F-|\vec{k}|)$,
\begin{eqnarray}
\Pi_{\omega \to V B N^{-1}}(\rho,P)&=&\int \frac{d^3k}{(2\pi)^3} \,
n(\vec{k}\,)~4~{t}_{\omega N \to \omega
N}(P^0+E_N(\vec{k}\,),\vec{P}+\vec{k}\,)\ ,
\label{eq:pid}
\end{eqnarray}
where $P^0,\vec{P}$ stand for the
energy and momentum of the $\omega$ in the nuclear
matter rest frame, and $E_N(\vec{k})=\sqrt{\vec{k}\,^2+M_N^2}$
is the nucleon energy.

\subsection{General remarks}
As one can see, we have followed here a relatively unconventional approach. Some
groups study the $\gamma N \rightarrow \omega N$, $\pi N \rightarrow \omega N$,
$\omega N \rightarrow \pi \pi N$ reactions incorporating a set of resonances,
their properties and couplings to $\gamma N$ or meson-baryon components being
fitted to data, using also constraints from better known reactions, such as
$\gamma N \rightarrow \pi N$, $\gamma N \rightarrow \pi \pi N$, $\pi N
\rightarrow  \pi \pi N$, etc. However, there are still large ambiguities, as
reflected from the different results obtained by different groups or even within
the same group \cite{Leupold:2009kz,Muhlich:2003tj,Muehlich:2006nn} . Other works like that of
Ref.~\cite{Lutz:2001mi} generate the resonances dynamically assuming that all
baryons other than $N$ and $\Delta(1232)$ are composite meson-baryon states. The
work has the advantatge of providing not only the resonances but also the
couplings, thus reducing uncertainties. Yet, it also has its limitations, since
not all resonances are dynamically generated, or at least some of them do not
qualify as meson-baryon quasibound states but rather as states made out of two
mesons and one baryon \cite{MartinezTorres:2007sr,Khemchandani:2008rk}. Some
also demand a subtle mixture of pseudoscalar-baryon and vector-baryon components
\cite{Garzon:2012np,Garzon:2013pad}.

\begin{figure}[htb]
\includegraphics[width=0.6\textwidth]{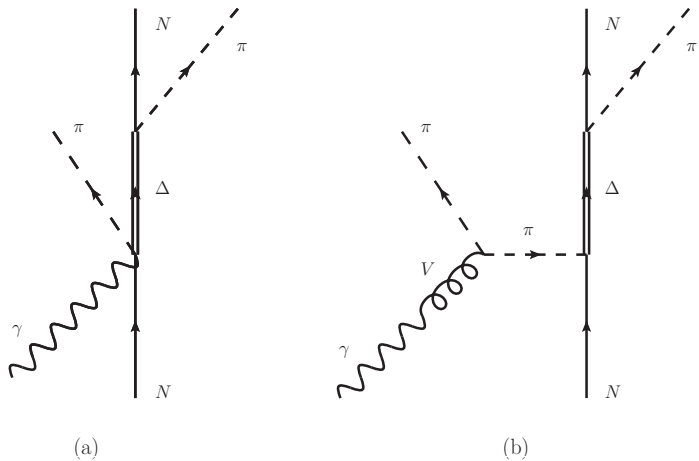}
\caption{ Most relevant mechanism for the $\gamma N \rightarrow  \pi \pi N$ reaction (a), and its
interpretation within the local hidden gauge approach (b).}%
\label{fig:fotoprod}%
\end{figure}

Arguments for a limited value of explicit analogies in related reactions can be found in the $\gamma N \rightarrow  \pi \pi N$ process. A successful model for this reaction was developed in \cite{GomezTejedor:1993bq,GomezTejedor:1995pe} where the $\Delta$ resonance played a dominant role in a diagram as that of Fig.~\ref{fig:fotoprod}(a), which in the local hidden gauge approach is interpreted as the diagram of Fig.~\ref{fig:fotoprod}(b) plus its associated Kroll Ruderman term \cite{Garzon:2013pad}. Note however that, while $V$ in diagram Fig.~\ref{fig:fotoprod}(b) stands in principle for the neutral vector mesons $\rho$, $\omega$, $\phi$, the $\omega$ meson does not contribute because of the G-parity forbidden $\omega \to \pi \pi$ decay, thus the link between the $\gamma N \rightarrow  \pi \pi N$  and $\omega N \rightarrow  \pi \pi N$ reactions is lost.

Our approach has looked in detail into the decay channels of the $\omega$ in the medium, including some which are overlooked in other works, as those studied in the two former subsections. The most important contribution, coming from $\omega \to \rho \pi$ decay, relies on an explicit calculation of the $\rho N$ scattering matrix which generates dynamically some resonances, as studied in \cite{Oset:2009vf, Garzon:2012np,Garzon:2013pad}.

\section{Results}
\label{sec:results}

We start this section by showing our results for the contributions to the
in-medium $\omega$ width coming from the $\omega \to 3\pi$ process. We recall
that, due to the coupling of the pion to $ph$ excitations in the medium, this
channel would mainly correspond to absorption processes of the type $\omega N
\to \pi
\pi N$, together with higher order contributions in density of the type  $\omega N N
\to \pi N N$.

In Table \ref{tab:width} we give the
in-medium width correction $\Delta\Gamma_{\omega\to\rho\pi}$ for a $\omega$
meson at rest in nuclear matter at normal saturation density employing two
different models for the $\omega \to \rho\pi$ interaction and dressing the
$\rho$ with the phenomenological parametrization of the width given by
Eq.~(\ref{eq:rhowidth}).
The first model
takes a $\omega \to \rho\pi$ coupling value of $G=11.9$ GeV$^{-1}$,
obtained from
a fit to vector meson radiative decays \cite{Gudino:2011ri}. Since this value
of $G$ does not give the full $3\pi$ width of a free $\omega$, the model must be
complemented by a contact term, as already explained in
Sect.~\ref{subsec:rhopi}. The contribution of these two terms to the in-medium
$\omega$ width correction are shown in Table
\ref{tab:width}, together with their sum. The
$\rho \pi$ mechanism represents about 60\% of the $3\pi$ strength. The second model corresponds to a totally
different strategy, which consists of adjusting the 
the $\omega \rho \pi$ coupling to a value $G=15.7$ GeV$^{-1}$, which reproduces
the full $\omega \to 3\pi$ free width.  Interestingly, the results obtained by the two
models are quite similar. This is a
welcome feature since any other model employing a different value of $G$, 
when complemented by the appropriate contact-term,
should produce values of the
in-medium $\omega$ width correction in between the ones obtained for the two
extreme models
considered in this work.

\begin{table}[htbp]
    \setlength{\tabcolsep}{0.3cm}
\begin{center}
%\begin{sidewaystable}[htbp]
\begin{tabular}{l|cc}
 &  $G=11.9$ GeV$^{-1}$ & $G=15.7$ GeV$^{-1}$ \\
\hline
$\omega \to \rho \pi$ & 54.5 & 94.9 \\
contact & 36.9 & -- \\
total $\Delta\Gamma_{\omega\to 3\pi}$ & 91.4 & 94.9
 \\
\hline
\end{tabular}
\end{center}
\caption{Contributions to the in-medium correction of the $\omega$ width $\Delta\Gamma_{\omega\to 3\pi} $ at $P^0=m_\omega$ and $\vec{P}=0$ in nuclear matter at  normal
saturation density, for two different $\omega\to \rho\pi$ coupling models, in
the case of dressing the
$\rho$ with the phenomenological parametrization of
Eq.~(\ref{eq:rhowidth}) and taking a
pion form-factor cut-off of $\Lambda_\pi=1200$ MeV.}
\label{tab:width}
\end{table}

Unless it is explicitly mentioned, the results shown in this section will be calculated using the effective coupling of $G=15.7$ GeV$^{-1}$. The results of Table \ref{tab:width2} intend to estimate
how the in-medium $\omega$ width correction coming from the $\omega \to \rho \pi$ mechanism depends on the particular in-medium properties of the $\pi$ and $\rho$ mesons. We present results for three different values of the pion
form-factor cut-off, $\Lambda_\pi=1000$, 1200, and 1400 MeV, which are the
appropriate
ones for
monopole type form factors employed in studies of the nucleon-nucleon
interaction \cite{Machleidt:1987hj}. The first row gives the results obtained when only the $\pi$ meson is dressed, while the results of the next rows incorporate the additional dressing of the $\rho$ meson according to the three different approaches considered in this work: a)
a simple parametrization consisting in an increase of the $\rho$ width by 33\% at $\rho=\rho_0$ (second row), b) the approach that mimicks the results of Ref.~\cite{Cabrera:2000dx} by implementing the coupling to an explicit $N^*(1520)$ resonance  (third row) and, finally, c) a $\rho$ self-energy obtained from a $t_{\rho N\to \rho N}$ amplitude derived entirely from the hidden gauge formalism that generates the $N^*(1520)$ dynamically (fourth row).
\begin{table}[htbp]
    \setlength{\tabcolsep}{0.3cm}
\begin{center}
%\begin{sidewaystable}[htbp]
\begin{tabular}{l|ccc}
 &  $\Lambda_\pi=1000$ MeV & $\Lambda_\pi=1200$ MeV &   $\Lambda_\pi=1400$ MeV \\
\hline
only $\pi$ & 69.0 & 76.5 & 79.2 \\
$\pi + \rho$(param) & 85.1 & 94.9 & 98.9\\
$\pi + \rho$($N^*N^{-1}$)  & 128.7 & 144.5 & 153.5 \\
$\pi + \rho$($t_{\rho N\to \rho N}$)  & 91.1& 101.2 & 106.4 \\
\hline
\end{tabular}
\end{center}
\caption{Contributions to the in-medium correction of the $\omega$ width $\Delta\Gamma_{\omega\to 3\pi} $ at $P^0=m_\omega$ and $\vec{P}=0$ in nuclear matter at  normal
saturation density, for three different values of the
pion form-factor cut-off and three different models for the $\rho$-meson self-energy.}
\label{tab:width2}
\end{table}
We observe that the in-medium pion self-energy represents the most important contribution  to the $\omega$ width, as already noted in \cite{Riek:2004kx}. Dressing the $\rho$ also adds a substantial contribution, more moderate in the case of the phenomenological model for the $\rho$ self-energy because the low-energy components of the $\rho$ spectral function play a significant role, as stressed recently in \cite{Cabrera:2013zga}, where they are referred to as "space-like" contributions. This also explains that the approach using an explicit $N^*$ component, with more strength at low energy, gives a larger value for the in-medium $\omega$ width correction, in qualitative agreement with what is found in \cite{Cabrera:2013zga}.
We have checked that, for a given model of the $\rho$ self-energy, changes in other
components of the pion self-energy besides the form-factor cut-off, such as employing a constant Landau-Migdal
parameter $g^\prime=0.6$, or implementing relativistic corrections at the $\pi N
N$ and $\pi N \Delta$ vertices, or using a static version of the pion form
factor, do not alter the results beyond the range of values shown in Table
~\ref{tab:width2}.

Adding the free decay width of the $\omega \to \rho \pi$ channel, 7.57~MeV, and dessing the $\rho$ meson with the $t_{\rho N\to \rho N}$ amplitude derived entirely from the hidden gauge lagrangians in a unitary coupled channel model, we obtain $\Gamma_{\omega}=109\pm 10$ MeV, the error being associated to variations in the parameters of the $\pi$ meson self-energy within reasonable limits.

We would like to comment on the fact that Eq.~(\ref{eq:self_rhopi3}) also
provides a real part of the $\omega$ self-energy in the medium, ${\rm
Re}\,\Pi_{\omega \to \rho
\pi}(\rho,P)$. When subtracting the free space contribution,
${\rm
Re}\,\Pi_{\omega \to \rho
\pi}(\rho=0,P)$, one could obtain an estimate of the $\omega$
mass shift. However, the calculation requires to perform integrals over
high momentum and energy components of the $\pi$ and $\rho$ propagators, on
which one does not have a good knowledge.
We have explored various strategies
and have obtained $\omega$ mass shifts ranging from $-100$ to $100$ MeV,
meaning that it is not possible to obtain a sensible stable result. Therefore
we focus only on discussing the medium modifications of the $\omega$ width. 
 
\begin{figure}[htb]
\includegraphics[width=0.6\textwidth]{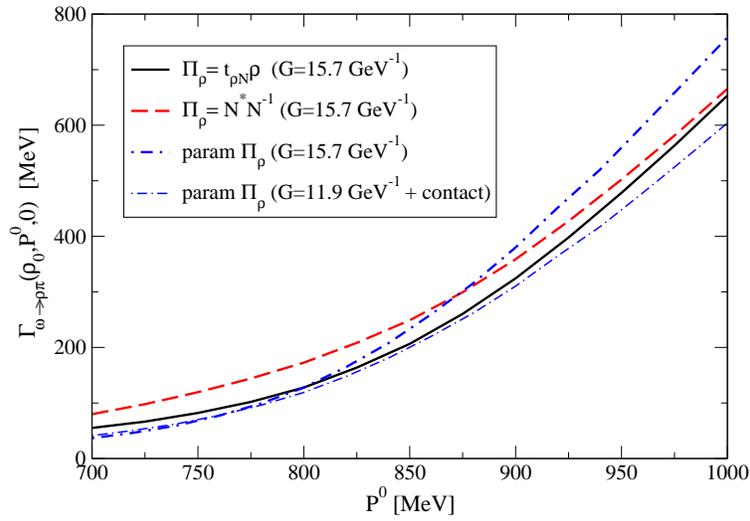}
\caption{Width of the $\omega$ meson at normal nuclear matter density and
$\vec{P}=0$ as a
function of its energy $P^0$. }%
\label{fig:width_energy}%
\end{figure}
\begin{figure}[htb] 
\includegraphics[width=0.6\textwidth]{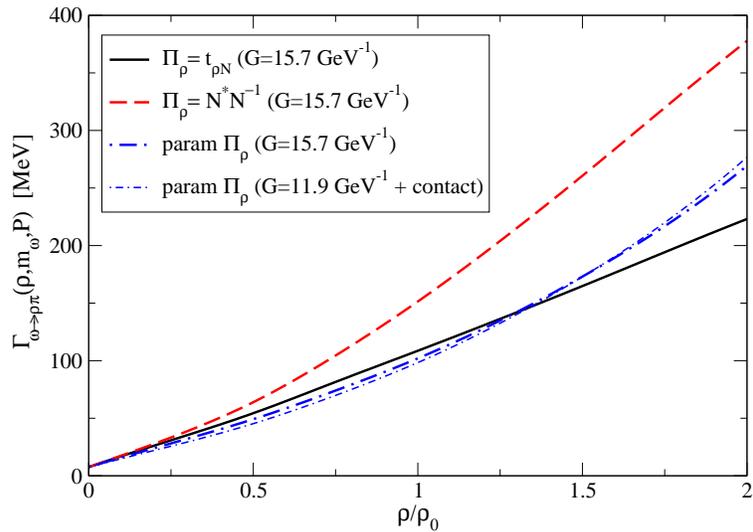}
\caption{Width of the omega meson at $P^0=m_\omega$ and $\vec{P}=0$ as a
function of
the nuclear matter density.}%
\label{fig:width_density}%
\end{figure}

In Fig.~\ref{fig:width_energy} we show
the width of a $\omega$ meson at rest
in nuclear matter at normal density as a function of the $\omega$
energy $P^0$, for the three prescriptions of the $\rho$-meson self-energy employed in this work and using the model with $G=15.7$ GeV$^{-1}$.  
The in-medium $\omega$ width increases smoothly with energy for all the
$\rho$-dressing models employed, the phenomenological one (thick dash-dotted
line)  presenting a stronger dependence due to the p-wave nature of the employed
parametrization, Eq.~(\ref{eq:rhowidth}). In this case, results are also shown
for the model that uses $G=11.9$ GeV$^{-1}$ plus a contact $\omega \to 3\pi$
term (thin dash-dotted line). We observe that, up to the free
$\omega$ mass, $m_\omega=783$~MeV, both models present a similar behavior but
beyond
this energy the width of the model employing $G=11.9$ GeV$^{-1}$ and a
contact-term contribution evolves more slowly, an effect tied to the
weaker energy dependence of the contact term.

The dependence of the $\omega$
width on the nuclear matter density is shown in Fig.~\ref{fig:width_density}, for the three prescriptions of the $\rho$-meson self-energy and employing the model with $G=15.7$ GeV$^{-1}$.
We observe a smooth increase of the $\omega$ width with density. The different non-linear density effects observed in the $\omega$ width are tied to the particular way the $\rho$-meson is dressed. In general, the models that have a larger $\rho$-meson strength at low energies also leave a stronger imprint on the $\omega$ self-energy, thus making the non-linear density effects to be more emphasized. An exception  occurs for the phenomenological parametrization of  Eq.~(\ref{eq:rhowidth}). In this case, the p-wave nature of the $\rho$ width makes it to increase more rapidly with density, hence magnifying the non-linear density effects of the $\omega$ width. For this approach to the $\rho$ self-energy, we also present in Fig.~\ref{fig:width_density} a calculation for the model that uses $G=11.9$ GeV$^{-1}$ and a
contact-term, obtaining essentially the same behavior with density as that for the $G=15.7$ GeV$^{-1}$ model.

From the low density behavior of the $\omega$ width, we can derive the imaginary part of the scattering length. This is done via the low-density theorem
\begin{eqnarray}
\Pi_{\omega} (\rho \rightarrow 0)= t_{\omega N\to \omega N} \rho \ .
\end{eqnarray}
Knowing
\begin{eqnarray}
a_{\omega N}=-\frac{1}{4 \pi}\frac{M_N}{M_N + m_\omega} t_{\omega N\to \omega
N} \ ,
\end{eqnarray}
with $t_{\omega N\to \omega N}$ being the amplitude at threshold,
and
\begin{eqnarray}
\Gamma_{\omega}=-\frac{{\rm Im}\, \Pi_{\omega}}{\omega_{\omega}} \ , 
\end{eqnarray}
we find
\begin{eqnarray}
{\rm Im}\, a_{\omega N}=\frac{1}{4 \pi} \frac{M_N}{M_N+m_{\omega}}
\frac{m_{\omega} \Gamma_{\omega}}{\rho} \ .
\end{eqnarray}
Taking our results for the $\omega$ width at $\rho=0.1 \rho_0$, in the case the
$\rho$ meson is dressed from the unitarized $t_{\rho N \to \rho N}$
amplitude, we obtain
\begin{eqnarray}
{\rm Im}\, a_{\omega N} \approx 0.39 \ {\rm fm} \,
\end{eqnarray}
which compares favourably with the values ${\rm Im}\, a_{\omega N}  \approx
0.31 \ {\rm fm}$ from \cite{Muehlich:2006nn}, ${\rm Im}\, a_{\omega N}  \approx
0.30 \ {\rm fm}$ from \cite{Klingl:1998zj} and ${\rm Im}\, a_{\omega N} 
\approx 0.20 \ {\rm fm}$ from \cite{Lutz:2001mi}. Not surprisingly our value is
somewhat bigger since our approach incorporates more decay channels.

\begin{figure}[htb]
\begin{center}
\includegraphics[width=0.6\textwidth]{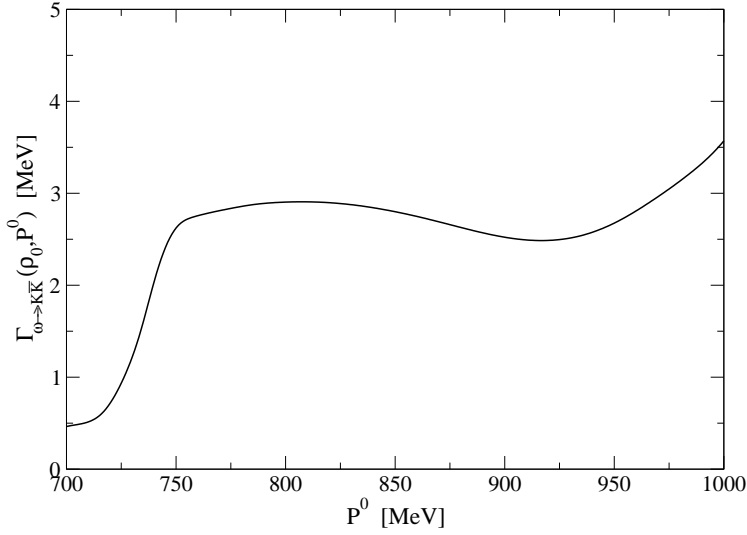}
\caption{ In-medium $\omega$ width coming from its coupling to $K\bar{K}$, at
$\rho_0=0.17 \ {\rm fm^{-3}}$  and
as a function of the $\omega$ energy $P^0$.}
\label{fig:4}
\end{center}
\end{figure}

We now present in Fig.~\ref{fig:4} the in-medium $\omega$
width correction coming from its coupling to
$K\bar{K}$ states and their in-medium excitations. We observe that this
contribution is very small compared with the values for the $\omega$
width coming from the anomalous decay
$\omega \rightarrow \rho \pi$ in nuclear matter. At normal nuclear matter
saturation density
and energies around the free $\omega$ mass, the in-medium width
correction associated to the $\omega \to K\bar{K}$ transition for a $\omega$
meson at rest is 
$\Delta\Gamma_{\omega \to K\bar{K}}(\rho_0,m_\omega)=-\mathrm{Im}\,\Pi_{\omega \to K\bar{K}}(\rho_0,m_{
\omega})/m_{\omega}=2.9$ MeV. This width is mainly coming from the
$p$-wave coupling of the antikaon to
$Yh$ components, since these are the lowest possible ${\bar K}$ excitations in
the medium. This
$\omega \to K{\bar K}$ in-medium width correction is therefore
associated to processes such as
 $\omega N\to K\Sigma$ and $K\Lambda$,  which are above their thresholds by $36$
MeV and $110$ MeV, respectively.
%, and higher density processes, intervening in $s$ wave, such as $\omega
% NN\to K\bar{K}NN\to K\pi\Lambda N\to K\Lambda N$. 
The antikaon-nucleon s-wave driven interaction terms, such as $\bar K N \to \pi
\Lambda,~\pi\Sigma$, which are significantly dominant over the p-wave $\bar K N
\to Y$ ones, do not contribute to the related $\omega  N\to
K\bar{K}N\to K\pi\Lambda, K\pi\Sigma$ processes, since their corresponding thresholds are closed by 28 MeV
and 102 MeV, respectively.

\begin{figure}[t]
\begin{center}
\includegraphics[width=0.6\textwidth]{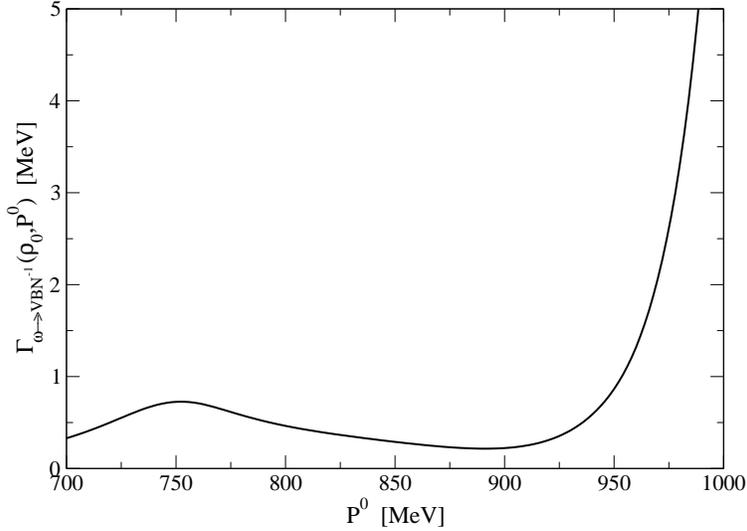}
\caption{Width of a zero momentum $\omega$ meson due to
the s-wave $\omega N \to VB$ interaction, at normal
nuclear matter density $\rho_0=0.17 \ {\rm fm^{-3}}$ and as a function of the
$\omega$ energy $P^0$.}
\label{fig:vb}
\end{center}
\end{figure}
We finally present in 
Fig.~\ref{fig:vb} the $\omega$ width correction associated
to the elastic an inelastic processes implemented by the $s$-wave interaction of
$\omega N$ with
vector mesons and baryons as a
function of
the $\omega$ energy. We observe that this contribution
produces a very small
$\omega$ width correction, $\Delta\Gamma_{\omega \to VB N^{-1}}(\rho_0,m_\omega) \sim 0.5$~MeV
for energies around the free
$\omega$ mass and at normal nuclear saturation density $\rho_0$. This result
would not change if a self-consistent calculation of the in-medium $t_{\omega N
\to \omega N}$ amplitude, implementing the
full width of the $\omega$ meson in the loop function of the $\omega N$
channel, was attempted. The same applies to a possible incorporation of the
in-medium width of the $\rho$ meson. This is due to the fact that the diagonal
$\omega N-\omega N$ and non-diagonal $\omega N - \rho N$ couplings
are zero \cite{Oset:2009vf} and any change in the $\omega$ or $\rho$ widths will only affect the
$\omega N-
\omega N$ amplitude via the indirect coupling to $K^* Y$ $(Y=\Lambda,\Sigma)$
states, lying about 300 MeV above the $\omega N$ threshold, in higher order
scattering processes. The small $\omega$ width correction obtained from the
$s$-wave interaction of the vector mesons and baryons
should be associated to the elastic $\omega N \to \omega N$ and the
inelastic $\omega N \to  \rho N$ processes, which are the only possible
decay channels
allowed in the employed vector-baryon interaction model
\cite{Oset:2009vf}. 

The implementation of pseudoscalar mesons into the scheme,
hence opening vector-baryon to pseudoscalar-baryon transitions such as $\omega N
\to \pi N$, does not seem to change the $t_{\omega N\to \omega N}$ amplitude significantly
\cite{Garzon:2012np}. We note that the coupling between the vector-baryon
and the pseudoscalar-baryon channels implemented by the box-diagrams
built up in \cite{Garzon:2012np} is done through t-channel pseudoscalar meson
exchange, while the anomalous t-channel vector-exchange terms are not
considered. Actually, a proper implementation in the $\omega \to \rho
\pi$ decay channel of the low energy branch of the $\rho$-meson self-energy,
associated to $ph$ excitations via a $\rho NN$-type coupling, would 
provide these type of terms, giving rise to an additional contribution to the $\omega$ width coming
from the $\omega N \to \pi N$ transition. However, our simple parametrization of
the $\rho$-meson self-energy, given in Eq.~(\ref{eq:rhowidth}), or the other two prescriptions miss this source of 
low energy strength of the $\rho$ spectral function. For this reason, we adopt
the pragmatic, safe and model independent point of view discussed by Friman in
\cite{Friman:1997ce}, where the contribution of the $\pi N$ channel to the width
of the $\omega$ meson was obtained from the use of detailed balance (to relate
the measured $\pi^- p \to \omega n$ cross section to the one of the reversed
process $\omega n \to \pi^- p$) and unitarity (to connect the value of this
cross section with the $\pi^- p$ contribution to the $\omega n$ forward
scattering amplitude). A fit to the $\pi^- p \to \omega n$ data provided a $\pi N$ channel contribution to the width of the $\omega$ meson of about
9 MeV \cite{Friman:1997ce}.

Joining all the contributions together, we conclude that the width of the
$\omega$ meson at rest in nuclear matter at normal saturation density is
$\Gamma_{\omega}(\rho_0,m_\omega)=7.6$~MeV (free width)$ +
108.7$~MeV ($\omega N \to \pi\pi N, \omega N N \to \pi NN$)$+ 2.9$~MeV
($\omega N \to K Y$)$+ 0.5$~MeV ($\omega N \to K^* Y \to \rho N$)$ +9$~MeV
($\omega N \to \pi N$)$= 129 \pm 10$ MeV, where the 10 MeV
error is associated to uncertainties of the
theoretical model. We
note that one
could add one more MeV to account for the other free decay channels of the
$\omega$ meson,  $\omega \to \pi^0 \gamma$ and $\omega \to \pi^+ \pi^-$.

Our calculated value of the width of the $\omega$ meson at rest in nuclear
matter is larger than that found
by other works in the literature. A few models derive the $\omega$ width from
the elastic $\omega N$ amplitude by means of the low-density theorem, finding
$\omega$ width values 
of the order of $40-60$~MeV \cite{Klingl:1998zj,Lutz:2001mi,Muehlich:2006nn} up
to 75~MeV \cite{Martell:2004gt}. The works of 
Refs.~\cite{Lutz:2001mi,Muehlich:2006nn} employ amplitudes that have been fitted
to a variety 
of elastic and
inelastic $\gamma N$ and $\pi N$ data, implementing resonance contributions and
rescattering effects in unitary models. The self-consistent approach of
Ref.~\cite{Riek:2004kx}, studying the influence of the in-matter
spectral function of the pion in the 
vector-meson self-energies, is the one closer to the methodology employed in
the present work for the contribution to the $\omega$ width from the
$\rho \pi$ mechanism. Yet, a value of about 60 MeV is quoted in
Ref.~\cite{Riek:2004kx} while our result from this $\omega \to \rho\pi$ channel
is of the order of 100 MeV. We have traced back the differences to the much
softer form-factor of gaussian type,
$F^{\rm  g}(\vec{q}\,)={\rm exp}(-\vec{q}\,^2/\Lambda^2)$ with
$\Lambda=440$ MeV, employed in Ref.~\cite{Riek:2004kx}, compared with
the monopole form-factor, $F^{\rm m}(\vec{q}\,^2) = (\Lambda_{\pi}^2 - m_{\pi}^2)
/
[\Lambda_{\pi}^2 - q^{0\,2} + \vec{q}\,^2 ]$ with
$\Lambda_{\pi}=1200$~MeV, employed in the present work and taken from $NN$
interaction studies \cite{Machleidt:1987hj}. The ratio $(F^{\rm m}/F^{\rm 
g})^2$ depends on the loop momentum over which one must integrate
to obtain the $\omega$ width but, for a characteristic value
of $\vec{q}=250$~MeV/c and setting $q^0=0$ MeV, it
amounts to 1.7, explaining the larger size of the $\omega$ width obtained in the
present work. 
Large values of the width are also reported in \cite{Cabrera:2013zga}, associated to the large modification of the $\rho$ spectral function at low invariant masses. We showed before that, when a microscopical model for the $t_{\rho N \to \rho N}$ amplitudes is used to evaluate the $\rho$ self-energy, this contribution is more moderate.

\begin{figure}[htb] 
\includegraphics[width=0.6\textwidth]{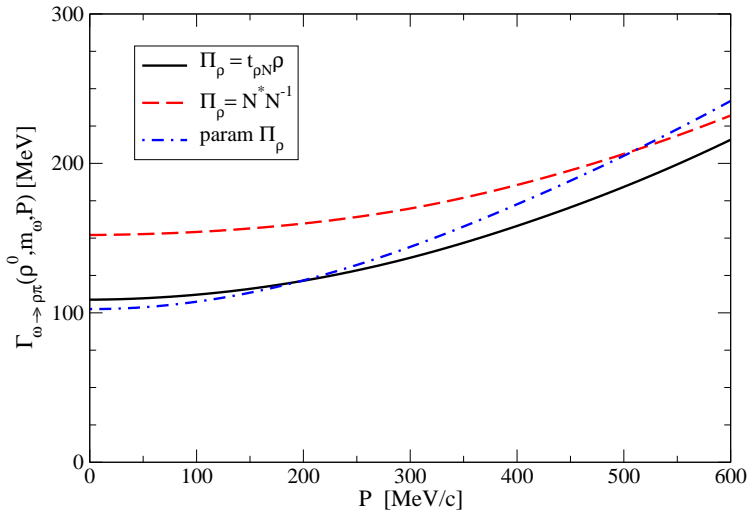}
\caption{Contribution to the $\omega$ width coming from the $\omega \to 3\pi$ channel in nuclear matter at saturation density, as a function of momentum, for the three different prescriptions of the $\rho$-meson self-energy employed in this work.
}%
\label{fig:width_mom}%
\end{figure}

In order to compare with the experimental determination of the $\omega$ width,
we need to extend our calculation to finite momentum. Although, in principle,
the $P^0$ and $\vec{P}$ variables of the $\omega$ meson are independent in
nuclear matter, one must correlate them on-shell if one wishes to compare with
the results of the $\omega$ in-medium width of \cite{Kotulla:2008aa}, since the
$\omega$ detection is done imposing the invariant mass of the $\pi^0 \gamma$
system to match the $\omega$ mass. In Fig.~\ref{fig:width_mom} we show the
momentum dependence of the $\omega$ width coming from the $\omega \to 3\pi$
channel in nuclear matter at saturation density, for the three different
prescriptions of the $\rho$-meson self-energy employed in this work. We observe
that $\Gamma_{\omega\to 3\pi}$ rises smoothly with momentum, and it can reach
values of about 200 MeV at $P=$ 600 MeV/c. We should keep in mind that we must
add about 15 MeV to these numbers from the other sources discussed above. The
experimental width is quoted to be $\Gamma_{\omega} \approx 130-150$ MeV for an
average 3-momentum of 1.1 GeV/c \cite{Kotulla:2008aa}. Apart from this global average, which is in qualitative agreement with what we obtain, the experimental analysis also reports values of $\Gamma_{\omega}$ for different momentum bins. One then sees that we obtain a good agreement within errors for the lower two momentum values reported in Fig.~ 4 of
Ref.~\cite{Kotulla:2008aa}, 400 MeV/c and 600 MeV/c, where our results should be more accurate.

Before closing this section, we note that the theoretical study of
\cite{Muehlich:2006nn} shows a somewhat stronger momentum dependence of the transverse
component of the $\omega$ self-energy and, consequently, the width evolves from
a
value of 60~MeV at zero momentum to around 130~MeV at 600~MeV/c, a value that is roughly consistent with the experimental results \cite{Kotulla:2008aa} . The new light brought into this problem by the present
work allows us to examine the results of \cite{Muehlich:2006nn} with a new
perspective. We recall that in that work the $\omega$ width is obtained in a
$t_{\omega N \to \omega N} \rho$ approximation from the elastic $\omega N$ amplitude,  which
is found to be
dominated by resonance hole components, the $1/2^-$ $S_{11}(1535)$ playing the
most important role. In the present work,
the main contribution to the $\omega$ width comes from the $\omega \to \rho
\pi$
channel, when the $\pi$ and $\rho$ mesons are allowed to be dressed in the
medium, which incorporates terms of higher orders in density. This  $\omega \to \rho
\pi$ process would
correspond to a background type term in \cite{Muehlich:2006nn}, claimed to be
less important than the resonance hole contributions. We believe that the
combination of a varied number of
resonances with a priori unknown couplings that are fitted to inelastic $\pi
N\to \omega N$ and
$\gamma N \to \omega N$ processes brings uncertainties to the
elastic $\omega N \to \omega N$ amplitudes, particularly when it comes to the
coupling to $\omega N$ states of a resonance like the $S_{11}(1535)$
which is quite below the $\omega N$ threshold. In our approach, the
resonant contributions to the $\omega N$ amplitude are generated from a unitary
model in coupled channels, including also $\rho N$, $\phi N$, $K^* \Lambda$ and
$K^* \Sigma$, and using the dynamics of the hidden gauge Lagrangians. The
$S_{11}(1535)$ is not obtained from the vector-baryon dynamics but it comes out
from the interaction of the pseudoscalar-baryon $\pi N$, $\eta N$, $K \Lambda$
and $K\Sigma$ channels \cite{inoue,Kaiser:1995cy}. One can consider
vector-baryon and pseudoscalar-baryon components together, as done in
\cite{Garzon:2012np}, but the mixing is small and the different dynamically
generated states basically keep their original identity. Yet, it is simple to
determine the coupling of one particular pseudoscalar-baryon resonance to a
vector-baryon channel using chiral unitary methods and vector meson dynamics, as
done in \cite{Doring:2008sv}. Neglecting terms of order ${\cal
O}\left((\vec{q}/2M)^2\right)$, the coupling of the $N^*(1535)$ to $\omega N$
states was found to be given by
\begin{equation}
-i\, t_{N^* N \omega} = g_{N^* N \omega} \, \vec{\sigma}\vec{\epsilon} \ ,
\label{eq:nstar}
\end{equation}
where $\vec{\epsilon}$ is the $\omega$ polarization (the zeroth component
$\epsilon^0$ was neglected since small momenta of the $\omega$ meson were being
considered). The dynamical model of Ref.~\cite{Doring:2008sv} determined
the value $g_{N^* N \omega} =0.02+i\, 0.28$. It is straightforward to show that,
neglecting terms of ${\cal O}\left((\vec{q}/2M)^2\right)$, the Lagrangian
employed in Ref.~\cite{Muehlich:2006nn},
\begin{equation}
{\cal L}_{1/2^- N \omega} = i  \bar{u}_R \gamma_5 \left( g_1 \gamma^\mu
\omega_\mu - \frac{g_2}{2 M_N} \sigma^{\mu\nu}\partial_\nu \omega_\mu \right)
u_N \ ,
\end{equation}
produces a term equivalent to that of
Eq.~(\ref{eq:nstar}),  as long as one identifies
 \begin{equation}
g_{N^* N \omega} = -g_1+g_2\frac{m_\omega}{2M_N} \ .
\end{equation}
Taking the values $g_1=3.79$ and $g_2=6.50$ quoted in \cite{Muehlich:2006nn},
one obtains $g_{N^* N \omega}=1.09$. Despite the considerable cancellation
between the $g_1$ and $g_2$ terms, this value is still much larger than the
absolute value of the coupling derived in \cite{Doring:2008sv}. Consequently,
the contribution of the $S_{11}(1535)$ to the $\omega$ width turns out to be
about 15 times larger in
\cite{Muehlich:2006nn} than that one would obtain using the coupling
derived in \cite{Doring:2008sv}. 
This is just a simple example trying to show that the couplings of
some resonances to $\omega N$ states at threshold might not be very well
constrained experimentally due to interferences between various a priori unknown
resonant terms and background contributions.

\section{Conclusions}
\label{sec:conclusions}

We have evaluated the width of the $\omega$ meson in the nuclear medium, from a
variety of processes.
 
We have first considered the free decay mode of the $\omega$ into three pions, which
is dominated by $\rho \pi$ decay, and, by replacing the $\rho$ and
$\pi$ propagators by their medium modified ones, we have obtained the medium
corrections to the $\omega$ width. In order to get a better feeling of uncertainties and place the results in  consistency with advances made recently on the theoretical description of the $\rho N$ interaction, 
we have
employed three different models for the $\rho$-meson propagator.
We have considered a phenomenological approach consisting in parameterizing the
$\rho$ spectral function by simply increasing the $\rho$ width by 33\% at normal nuclear matter density. We have also taken a more realistic approach that implements the coupling of the $\rho$ to an explicit $N^*(1520) h$ excitation, hence increasing the $\rho$ strength at low invariant masses. Finally, we have implemented a recently derived $\rho N$ scattering amplitude, based on coupled-channel unitarizing techniques using local hidden gauge Lagrangians.
The results of the three approaches are qualitatively similar, but our analysis is useful to show the relevance of the role of the resonances like the $N^*(1520)$, which are dynamically generated in the latter approach. We establish the relevance of this resonance on the in-medium width of the $\omega$ meson and shed light on results obtained with approaches that incorporate it more phenomenologically.
We have also taken two
different prescriptions for the
$\omega\rho\pi$ coupling $G$, a value adjusted to
reproduce vector meson
radiative decays, in which case the model must be supplemented by the additional
consideration of a contact term, or an effective coupling  adjusted to reproduce the
free decay width of the $\omega$ meson into three pions. Both prescriptions give rise to very similar results for the in-medium $\omega$ width.

 We have also taken into account the
contributions from the virtual $\omega \to K \bar K$ decay channel, which is open in
the medium due to the excitation of $Yh$ components with antikaon quantum
numbers. This $\omega$ decay channel has provided a very small contribution to
the $\omega$ decay width, less than 3 MeV, because the dominant s-wave
components of the $\bar K N$ interaction, leading to $\omega N \to K Y \pi$
transitions, are not energetically allowed. The only possible transitions are
$\omega N \to K Y$, which are triggered by the much weaker p-wave component of
the $\bar K N$ interaction at the low energies explored by the processes studied
here. 

Finally, we have obtained the quasielastic and inelastic mechanisms induced by a
$t_{\omega N \to \omega N} \rho$ term, where $t_{\omega N\to \omega N}$ is the $\omega N$ scattering
matrix evaluated within a unitary scheme in coupled channels of vector-baryon
type. This gives rise to a practically negligible contribution, of 0.5 MeV, due
to the  zero value of the tree-level $\omega N - \omega N$ and $\omega N - \rho N$
amplitudes of the employed model. Unitarization through the coupling to $K^*Y$
channels gives a non-zero but small value for these amplitudes. The extension of
the model to incorporate the pseudoscalar-baryon channels through box diagrams
employing normal $VPP$ couplings does not change the situation significantly. We have
therefore added a 9 MeV contribution to the $\omega$ width from the $\omega N\to
\pi N$ processes, following a model independent approach based on detailed
balance and unitarity. % \cite{Friman:1997ce}.

With all these contributions, we obtain a value of $129\pm 10$
MeV for the width of the
$\omega$ meson at rest in nuclear matter at normal saturation density, which is
a substantially larger width than that quoted in previous theoretical works, many of them based on an indirect extraction of the $\omega N$ scattering amplitude from fits to  $\gamma N \rightarrow \omega N$, $\pi N \rightarrow \omega N$, $\omega N \rightarrow \pi \pi N$ data.

We have also evaluated the momentum dependence of the in-medium $\omega$ width and show that it increases with momentum, being about 200 MeV at 600 MeV/c. The range of values that we obtain up to 600 MeV/c are in line with the width of the order of 130-150 MeV obtained by the CBELSA/TAPS collaboration in photoproduction reactions off nuclei for an average momentum of around 1.1 GeV/c. More specifically, we obtain a very good agreement, within errors, with the $\Gamma_{\omega}$ results reported for the two lower momentum bins of 400 MeV/c and 600 MeV/c.

\section*{Acknowledgments}  
We would like to acknowledge useful discussions with U. Mosel.
This work is partly supported by the Spanish Ministerio de Economia y Competitividad and European FEDER funds under the contract numbers
FIS2011-28853-C02-01, FIS2011-24154, the Generalitat Valenciana in the program Prometeo, 2009/090 and
Grant No. 2009SGR-1289 from Generalitat de Catalunya. L.T. acknowledges support from Ramon y Cajal Research Programme, and from FP7-PEOPLE-2011-CIG under contract PCIG09-GA-2011-291679. We acknowledge the support of the European Community-Research Infrastructure
Integrating Activity
Study of Strongly Interacting Matter (acronym HadronPhysics3, Grant Agreement
n. 283286)
under the Seventh Framework Programme of EU.

\end{document}